\documentclass[a4paper,11pt]{article}
\usepackage[DIV10]{typearea}
\usepackage[latin1]{inputenc}
\usepackage[T1]{fontenc}
\usepackage{graphicx}
\usepackage[english]{babel}
\usepackage{amssymb}
\usepackage{geometry}
\usepackage{color}
\usepackage{xspace}
\usepackage{textcomp}
\usepackage{hyperref}
\usepackage{epic}
\usepackage{cancel}
\usepackage{amsmath,bm}
\usepackage{epstopdf}
\usepackage{ifthen}
\usepackage[footnotesize]{caption2}

\DeclareMathSymbol{\Gamma}{\mathalpha}{letters}{"00}
\DeclareMathSymbol{\Delta}{\mathalpha}{letters}{"01}
\DeclareMathSymbol{\Theta}{\mathalpha}{letters}{"02}
\DeclareMathSymbol{\Lambda}{\mathalpha}{letters}{"03}
\DeclareMathSymbol{\Xi}{\mathalpha}{letters}{"04}
\DeclareMathSymbol{\Pi}{\mathalpha}{letters}{"05}
\DeclareMathSymbol{\Upsilon}{\mathalpha}{letters}{"07}
\DeclareMathSymbol{\Phi}{\mathalpha}{letters}{"08}
\DeclareMathSymbol{\Psi}{\mathalpha}{letters}{"09}
\DeclareMathSymbol{\Omega}{\mathalpha}{letters}{"0A}
\DeclareMathSymbol{\varGamma}{\mathalpha}{operators}{"00}
\DeclareMathSymbol{\varDelta}{\mathalpha}{operators}{"01}
\DeclareMathSymbol{\varTheta}{\mathalpha}{operators}{"02}
\DeclareMathSymbol{\varLambda}{\mathalpha}{operators}{"03}
\DeclareMathSymbol{\varXi}{\mathalpha}{operators}{"04}
\DeclareMathSymbol{\varPi}{\mathalpha}{operators}{"05}
\DeclareMathSymbol{\varSigma}{\mathalpha}{operators}{"06}
\DeclareMathSymbol{\varUpsilon}{\mathalpha}{operators}{"07}
\DeclareMathSymbol{\varPhi}{\mathalpha}{operators}{"08}
\DeclareMathSymbol{\varPsi}{\mathalpha}{operators}{"09}
\DeclareMathSymbol{\varOmega}{\mathalpha}{operators}{"0A}

\newcommand{\D}{\mathrm{d}}

\def\beq{\begin{equation}}
\def\eeq{\end{equation}}
\def\bea{\begin{eqnarray*}}
\def\eea{\end{eqnarray*}}

\newcommand{\stau}{{\widetilde{\tau}}}
\newcommand{\sTop}{{\widetilde{t}}}	

\newcommand{\mstau}{m_{\stau_1}}
   
\newcommand{\mne}{{m_{\s{\chi}^0}}}
  
\newcommand{\msq}{{m_{\widetilde{q}}}}
\newcommand{\mgo}{{m_{\widetilde{g}}}}
\newcommand{\mlcp}{{m_{\text{LCP}}}}
\newcommand{\sq}{{\widetilde{q}}}
\newcommand{\go}{{\widetilde{g}}}
\newcommand{\neu}{{\widetilde{\chi}^0}}

\newcommand{\mstop}{m_{\s{t}_1}}
\newcommand{\pt}{p_{\text{T}}}

\newcommand{\R}{r_{\beta}}	

\newcommand{\GEV}{\ensuremath{\,\textnormal{GeV}}}
\newcommand{\TEV}{\ensuremath{\,\textnormal{TeV}}}

\newcommand{\pb}{\ensuremath{\,\textnormal{pb}}}

\newcommand{\ifb}{\ensuremath{\,\textnormal{fb}^{-1}}}

\newcommand{\s}[1]{\widetilde{#1}}

\newcommand{\ME}{{\textsc{MadEvent}}}
\newcommand{\PY}{{\textsc{Pythia}}}
\newcommand{\DP}{{\textsc{Delphes}}}
\newcommand{\WZ}{{\textsc{Whizard}}}

\newcommand{\IREL}{I_{\text{rel}}}

\definecolor{evgray}{gray}{.48}
\definecolor{orange}{RGB}{255,127,0}
\definecolor{purple}{RGB}{147,112,219}
\definecolor{comjoern}{RGB}{130,5,255}						
\definecolor{comjan}{RGB}{255,40,30}								
\definecolor{gray}{RGB}{125,125,125}										
\definecolor{jnew}{RGB}{230,120,30}	
\definecolor{dgreen}{RGB}{0,180,10}

\begin{document}

\date{\mbox{ }}

\title{ 
{\normalsize  
6th March 2012 \hfill\mbox{}\\}
\vspace{2cm}
\bf
Long-lived staus from strong production
in a simplified model approach\\[8mm]}
\author{Jan Heisig and J\"{o}rn Kersten\\[2mm]
{\small\it II.~Institute for Theoretical Physics, University of Hamburg,
Germany}\\
{\small\tt jan.heisig@desy.de, joern.kersten@desy.de}
}

\maketitle

\thispagestyle{empty}

\vspace{1cm}

\begin{abstract}
\noindent
We study the phenomenology of a supersymmetric scenario where the next-to-lightest
superparticle is the lighter stau and is long-lived due to a very weakly coupled 
lightest superparticle, such as the gravitino. 
We investigate the LHC sensitivity and its dependence on the superparticle spectrum 
with an emphasis on strong production and decay. 
We do not assume any high-scale model for SUSY breaking but work along the lines 
of simplified models. 
Devising cuts that yield a large detection efficiency in the whole parameter space,
we determine the LHC's discovery and exclusion potential.
This allows us to derive robust limits on $\mstau$, $\mgo$, a common $\msq$, and 
$\mstop$. We briefly discuss the prospects for observing stopped staus.
\end{abstract}

\clearpage

\tableofcontents

\section{Introduction}

In supersymmetric scenarios where the gravitino is the lightest superparticle (LSP), 
the decay of the next-to-LSP (NLSP) is suppressed by the very weak gravitino 
interactions, as long as $R$ parity is conserved.  For a gravitino mass above a few 
keV, NLSPs produced at a collider usually have a decay length that is large compared
to the size of the detector. 
The same is possible in axino LSP scenarios. For a charged NLSP such as a stau, this 
leads to the exciting possibility of finding SUSY in events with no missing energy and 
two charged tracks leaving the detector. 

In a collider experiment such a long-lived stau can either be produced directly or in a 
decay chain following the initial production of a pair of superparticles. 
At the LHC the production of the strongly interacting squarks and gluinos clearly has the
biggest potential to dominate over other production mechanisms. This is why in this work 
we concentrate on the strong production.
Unfortunately, a considerable part of the more than 100 free parameters of the MSSM
influences the signature of staus via the appearance of intermediate sparticles in the
cascades.  
This general problem of the large SUSY parameter space has often been tackled by 
studying constrained models such as the CMSSM\@.
The nonobservation of SUSY has severely reduced the constrained model parameter 
space. Among other things, this has driven the interest in model-independent studies, 
including regions in parameter space that are not covered by constrained models. 

There are basically two ways of going beyond constrained models. The first is to use a 
well-motivated set of 19 free parameters in the framework of the phenomenological 
MSSM~\cite{Djouadi:1998di} and perform a Monte Carlo scan over the (still vast) 
parameter space, displaying the behavior of observables in a scatter plot (see 
e.g.\ \cite{Berger:2008cq}).
The second is to reduce the parameter space in a bottom-up approach more drastically 
after identifying the most important low-scale parameters that determine the signature.
This latter approach finds its realization in the so-called simplified models 
\cite{Alwall:2008ag,Alves:2011wf}.
So far the idea of simplified models has not been considered in the case of a very 
weakly interacting LSP\@. Nonetheless, it is especially suitable in the case of a 
long-lived stau scenario, as we will show in this paper.

We will determine the LHC sensitivity for a general long-lived stau\footnote{%
The analysis is virtually identical for any charged slepton NLSP\@.}
scenario utilizing a simplified-model approach. We consider the $8\TEV$ and 
the $14\TEV$ LHC runs. If not stated otherwise, we refer to the $8\TEV$ run.
We consider the particles of the MSSM as the only particles involved in the 
interactions inside the collider. 
We shall assume that there is no accidental phase space suppression that renders 
any sparticle other than the stau NLSP long-lived.

We will define a set of simplified models in section \ref{sec:simplified}. 
We will divide the problem into two parts---the production and the decay. 
We will first discuss how the production cross section depends on the
mass pattern in the squark and gluino sector. This discussion is general 
and applies to any LSP scenario. 
However, we will consider a range of sparticle masses around the LHC 
limits that are typically reachable in the long-lived stau scenario.
Our results will be presented for the two limiting cases of a common 
squark mass and of a scenario in which one of the stops is much lighter 
than all other squarks. In each case we will allow for two free parameters:
the gluino mass and either the common squark mass or the stop mass. 
We will then consider the 
decay of colored sparticles into the stau NLSP and
discuss the impact of the intermediate sparticles in the 
decay chain on the observables. This discussion
is based on the observation that the direct signature of the
stau (rather than SM particles from the cascade) provides the
most significant contribution to a potential discovery or an exclusion.
Since the identifiability of the stau depends
strongly on its velocity, our considerations are driven by the
quest of finding the quantities the stau velocity dominantly 
depends on. This necessitates introducing the stau mass as the
third free parameter. We will define three simplified models that 
serve as limiting cases and thus are able to
capture the phenomenology of any realistic spectrum within the 
long-lived stau scenario.

In section \ref{sec:bkgsel} we will study the relevant background sources
and discuss the rejection obtained by a cut on the velocity of a
stau candidate. Due to two distinct measurements of the 
velocity and the requirement of two staus per event,
an excellent background rejection is possible. For certain regions
in parameter space a large number of staus are likely
to be missed by the current trigger settings.  Hence, we will propose
a dedicated trigger in order to be able to record the
corresponding events.
In section \ref{sec:selection} we will introduce selection criteria
that achieve high efficiencies in the whole parameter space.

In section \ref{sec:expl} we will present the results obtained in
a Monte Carlo study by scanning over the defined parameter space.
We will show the discovery and exclusion reach in terms of
the stau, squark and gluino masses, both with and without utilizing the
proposed trigger setup.  These results represent conservative,
model-independent bounds and thus can be applied to all models with a
long-lived charged slepton.

Some staus may be stopped inside the detector and decay much later.
We will discuss the potential for observing such decays in section 
\ref{sec:stopped}, estimating the number of staus that are available for the 
analysis.

\section{Simplified models for long-lived stau scenarios}
\label{sec:simplified}

The task of simplified models is to reduce the huge SUSY parameter space.
The same is true for constrained models, but those obtain a reduction by
imposing boundary conditions with few parameters at a very high energy
scale.  In simplified models the reduction is driven by the signatures
at colliders.  We aim to determine the most important low-energy
parameters governing the LHC sensitivity to long-lived staus
originating from cascades following the strong production of SUSY
particles.

\subsection{Production} \label{sec:prod}

Obviously, the masses of the produced squarks and gluinos play an important
role---the production cross section depends on them. The first simplifying
assumption we discuss is a common mass for all squarks.
In the high-mass region, which we are primarily interested in, the
production of squarks requires partons with a large momentum fraction.
The corresponding parton distribution functions (PDFs) of $u$ and $d$
quarks are much larger than those of heavier quarks and of antiquarks.
Hence, governed by the large contribution of the first-generation squarks,
$\sq\sq$ production dominates over $\sq\overline\sq$ (see upper
right panel of figure~\ref{fig:scale}), and $\go\sq$ dominates over
$\go\overline\sq$. The latter is even negligible and thus not considered
here at all.

\begin{figure}
\centering
\setlength{\unitlength}{1\textwidth}
\begin{picture}(0.85,0.3)
\put(0.0,0.0){
  \put(0.04,0.035){\includegraphics[scale=1.1]{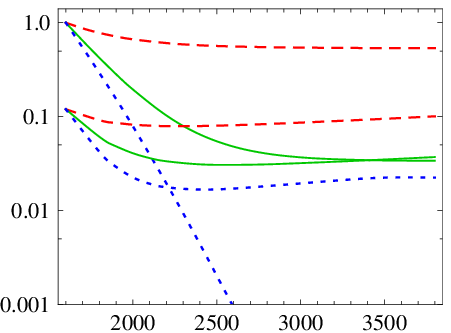}}
  \put(0.183,0.0){\footnotesize $m_i\; [ \GEV \, ] $}
  \put(0.0,0.131){\rotatebox{90}{\footnotesize $\sigma_i/\sigma^{\sq\sq}_{\text{deg}}$}}
  \put(0.32,0.219){\tiny $i= \s d$}
  \put(0.3,0.172){\tiny $i= \s u$}
  \put(0.19,0.126){\tiny $i= \s u,\s d$}
  \put(0.125,0.242){\tiny $\sq\sq$}
  \put(0.093,0.16){\tiny $\sq\overline\sq$}
  }
 \put(0.43,0){
  \put(0.04,0.035){\includegraphics[scale=1.1]{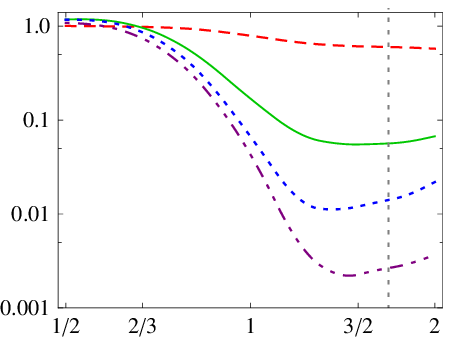}}   
  \put(0.197,0.0){\footnotesize $\mgo/\msq$}
  \put(0.0,0.131){\rotatebox{90}{\footnotesize $\sigma^{\text{full}}_i/\sigma^{\text{full}}_{\text{deg}}$}}
  \put(0.27,0.24){\tiny $\s d$}
  \put(0.27,0.198){\tiny $ \s u$}
  \put(0.298,0.145){\tiny $ \s u,\s d$}
  \put(0.21,0.076){\tiny all but $\sTop_1$}
  }
\end{picture}
\caption{NLO cross sections for the production of colored sparticles 
at the $8\TEV$ LHC\@.
\emph{Left:} Squark-squark (curves originating at 1 on the ordinate) and
squark-antisquark production (curves originating below) as a function of
the mass $m_i$ of different squark species.  All other squark 
masses are degenerate at $\msq=1600\GEV$ and $\mgo/\msq \simeq 1.68$.
The curves are normalized to the
squark-squark production cross section for the fully degenerate case,
i.e., $m_i=\msq$.
\emph{Right:}~Total cross section for all production channels (normalized
to the cross section with degenerate masses) along a typical LHC
sensitivity limit (see figure~\ref{fig:scale}) parametrized by the ratio
$\mgo/\msq$. We decoupled $\s d$, $\s u$, and both flavors, obtaining
the red dashed, green solid, and blue dotted lines, respectively.  The
purple dot-dot-dashed curve shows the cross section for $\s q = \s t_1$
and all other squarks decoupled. 
}
\label{fig:msqvar}
\end{figure}

The effects of abandoning the assumption of a common squark mass are
displayed in figure~\ref{fig:msqvar}.  The left panel shows 
the next-to-leading order (NLO) cross section for $\sq\sq$ production
computed by \textsc{Prospino~2} \cite{1997NuPhB.492...51B} as a function
of $m_{\s d}$ (red dashed curve originating at $1$ on the ordinate),
$m_{\s u}$ (green solid curve) and
$m_{\s u,\,\s d} = m_{\s d}=m_{\s u}$ (blue dotted curve).  In each
case, all other squark masses are degenerate at $\msq=1600\GEV$.  We
normalized the curves to the $\sq\sq$ production cross section at the
point where \emph{all} squarks are mass degenerate.  The ratio
$\mgo/\msq$ is chosen to be 1.68.  This is the value where $\sq\sq$
production contributes maximally to the total cross section of colored
sparticles in figure~\ref{fig:scale}. From figure~\ref{fig:msqvar}
it is obvious that the contribution of the first-generation squarks
is dominant---in the case where $\s d$ and $\s u$ masses increase 
simultaneously, the $\sq\sq$ cross section drops drastically. 
Decoupling either only $\s u$ or only $\s d$ has a much less drastic
effect. Decoupling the squarks other than $\s u$ and $\s d$ does not
significantly change the production cross section.
Additionally, we plot the analogous curves for $\sq\overline\sq$ production 
(curves originating at slightly above 0.1 on the ordinate). Here, the 
$t$-channel contribution, which introduces the flavor dependence, is much 
less important.  Consequently, if we decouple both
$\s d$ and $\s u$, the $\sq\overline\sq$ cross section drops much more
moderately than the one for $\sq\sq$ and hence becomes dominant.
As we stated above, $\go\overline\sq$ is negligible compared to
the other contributions, and so is $\go\sq$ in the case of decoupled
$\s u$ and $\s d$---the cross section drops by two orders of magnitude. 
Consequently, for decoupled first-generation squarks, 
$\sq\overline\sq$ and $\go\go$ remain the most important
production channels.  The latter channel is evidently much
less sensitive to the squark masses.

In the right panel of figure~\ref{fig:msqvar} we plot the total cross
section for all production channels
($\go\go$, $\go\sq$, $\sq\sq$, $\sq\overline\sq$) along an expected $8\TEV$
LHC sensitivity limit (black dashed line in the left panel 
of figure~\ref{fig:scale}) parametrized by the ratio $\mgo/\msq$.  As
before, different squark species are decoupled while the remaining
squarks have a common mass $\msq$, and for each value of
$\mgo/\msq$ the cross section is normalized to the cross section with
degenerate squark masses.  The vertical dotted line
marks the ratio $\mgo/\msq \simeq 1.68$ used in the left panel.  Stop
production is not included in the curves corresponding to decoupling
$\s d$ or $\s u$.  However, it would change only the blue dotted curve
noticeably, giving an enhancement by a factor $4/3$.  
In each case considered so far, the right- and left-handed squarks 
were treated uniformly.  Since the strong interaction is not
chiral, decoupling only
$\s q_\text{L}$ or $\s q_\text{R}$ merely results in a combinatorial
factor that does not involve information from the PDFs.

The lowermost curve in the right panel of figure~\ref{fig:msqvar}
corresponds to decoupling all squarks but the lighter stop.  Decoupling
all squarks except either $\s{t}_2$ or one $\s b$, $\s c$ or $\s s$
squark would give the same result.  This scenario provides the limiting
case of a minimal cross section.

We are left with two limiting setups, the one with a common squark mass
and the light stop scenario. Each contains two free parameters,
$\mgo$ and $\msq$, or $\mgo$ and $\mstop$, respectively. We will
focus on the former case but come back to the latter scenario
in section~\ref{sec:lightstop}. From the discussion in this 
section the LHC reach for a general scenario can be estimated.

\subsection{Decay}\label{sec:decay}

After the production of squarks and gluinos, these particles decay via a
cascade to the stau NLSP\@.  The mass difference between the lightest colored
sparticle (LCP) and the stau NLSP determines the total phase space 
available in the cascade. Thus, it strongly affects the kinematics 
of the stau (and of the SM particle radiation). We will therefore consider
as a third free parameter the stau mass.

In order to study the impact of the intermediate sparticles in the
cascade, we consider limiting cases.  Following the considerations in
\cite{Konar:2010bi,Horn:2009zx} and translating them into the stau NLSP
scenario we find that within the MSSM there are no spectra for
which the LCP preferably decays via chains containing more than three
intermediate sparticles between the LCP and NLSP\@.  In large regions of 
the parameter space, shorter decay chains give the dominant contribution.

We focus on the impact of the mass spectrum on the stau
velocity, which is the most important quantity affecting the identification 
of staus.  We consider the limit where only massless SM particles are
produced in the cascade.  For a two-body decay the velocity of the
daughter sparticle $i+1$ in the rest frame of the mother sparticle~$i$ is 
\beq
\beta^{(i)}_{i+1}=\frac{m_i^2-m_{i+1}^2}{m_i^2+m_{i+1}^2}\,.
\label{eq:betam}
\eeq
If the mother sparticle has velocity $\beta_i$ in the lab frame,
the velocity of the daughter sparticle in the lab frame reads
\begin{equation}
\beta_{i+1}=\sqrt{1-\frac{(1-\beta_i^2)\bigl(1-\beta^{(i)}_{i+1}{}^2\bigr)}{\bigl(1+\beta_i\,\beta^{(i)}_{i+1}\cos\theta^{(i)}_{i+1}\bigr)^2}}\,,
\label{eq:velotrans}
\end{equation}
where $\theta^{(i)}_{i+1}$ is the decay angle in the rest frame of 
the mother sparticle.
Assuming a fixed mass gap $m_0-m_n$ ($=\mlcp-\mstau$)
and considering an $(n-1)$-step decay%
\footnote{Following \cite{Alves:2011wf}, we refer to a cascade with $n$
intermediate sparticles between the LCP and the NLSP as an `$n$-step
decay'.
For instance, the decay $\sq\to\neu\to\stau$ is a $1$-step decay.
}
($m_0\ge m_1\ge \dots\ge m_n$)
with a uniform probability distribution for $\theta^{(i)}_{i+1}$ in each
decay (i.e.,
ignoring spin correlations), we can compute the mean of the NLSP velocity
$\beta_n$ as a function of all masses and the LCP velocity 
$\beta_0$,
\beq
\overline \beta_n=\overline\beta_n (\beta_0,m_1,\dots,m_{n-1}) \,.
\eeq
It turns out that $\overline \beta_n$ has one minimum (maximum) at the point
given by the mass pattern\footnote{We checked this explicitly
up to $n=3$ but expect it to hold for any $n$.}
\beq
\label{eq:masspatt}
m_i\simeq m_0^{\frac{n-i}{n}}m_n^{\,\frac{i}{n}}
\eeq
and $n$ maxima (minima) at
\begin{equation}
\label{eq:massdeg}
m_i=m_0\,,\quad m_j=m_n \quad 
\forall\; i<k,\; j\geq k\,, \quad k = 1,\dots,n \,.
\end{equation}
The extrema (\ref{eq:massdeg}) represent
the mass-degenerate limit and correspond
(in the approximation we are currently working in) to the direct 
decay of the LCP into the NLSP\@.
The result (\ref{eq:masspatt}) is not surprising: it
renders all velocities $\beta^{(i)}_{i+1}$ to be equal---on average
each decay gives the same contribution to the velocity of the NLSP\@.
In contrast, in (\ref{eq:massdeg}) one decay dominates over the others.
The result also implies that the extremal values of $\overline \beta_m$
lie between those of $\overline \beta_n$ if $m<n$. The $m$-step
decays represent a slice in the space of the masses in the $n$-step
decays with $n-m$ masses degenerate. This slice clearly does not
contain the point (\ref{eq:masspatt}).

\begin{figure}
\centering
\setlength{\unitlength}{1\textwidth}
\begin{picture}(0.85,0.4)
\put(0.0,0.0){
  \put(0.04,0.035){\includegraphics[scale=1.1]{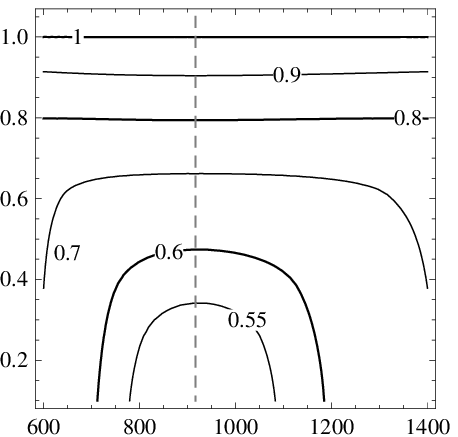}}
  \put(0.18,0.0){\footnotesize $m_1\; [ \,\text{a.u.} \, ] $}
  \put(0.0,0.198){\rotatebox{90}{\footnotesize $\beta_0$}}
  }
 \put(0.43,0){
  \put(0.04,0.035){\includegraphics[scale=1.14]{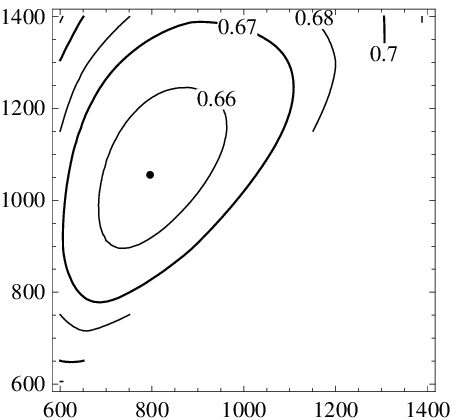}}   
  \put(0.18,0.0){\footnotesize $m_2\; [ \,\text{a.u.} \, ] $}
  \put(0.0,0.15){\rotatebox{90}{\footnotesize $m_1\;[ \,\text{a.u.} \, ]$}}
  }
\end{picture}
\caption{Contours of constant average velocity 
$\overline \beta_n(\beta_0,m_1,\dots,m_{n-1})$
for $m_0=1400$ and $m_n=600$ in arbitrary units. (The overall mass scale is
irrelevant. For convenience we chose values that could be realistic masses in GeV.)
\emph{Left:}~$\overline \beta_2$ as a function of $\beta_0$ and $m_1$.
The dashed vertical line denotes the extremum according to 
(\ref{eq:masspatt}). For $\beta_0\lesssim0.77$ it
is a minimum. Above this value it turns into a maximum, which is
not very pronounced, though, as $\overline \beta_2$
is nearly independent of $m_1$. 
\emph{Right:}~$\overline \beta_3$ as a function of $m_1$ and $m_2$
for $\beta_0=0.6$. The central dot denotes the minimum according to 
(\ref{eq:masspatt}). The maxima according to (\ref{eq:massdeg}) are
located in the lower-left, upper-left and upper-right corners.
The limiting case of an effective $1$-step decay chain ($n=2$) lies on
the borders of the contour plot---the upper border ($m_1=m_0$), the left
border ($m_2=m_3$) and the diagonal ($m_1=m_2$).
}
\label{fig:nmin}
\end{figure}

Whether (\ref{eq:masspatt}) is a minimum or a maximum depends on
$\beta_0$. In fact, (\ref{eq:masspatt}) is a maximum 
only if $\beta_0$ is very close to the speed of light (see the left panel
of figure~\ref{fig:nmin}). The high efficiency obtained in the long-lived stau 
search (see section \ref{sec:selection}) pushes the boundaries
of the LHC sensitivity to high squark and gluino masses. 
Hence, they will
typically be produced rather close to threshold and thus  $\beta_0$ 
is expected to be significantly below $1$, at least if LCP production
dominates.  In this case
\eqref{eq:masspatt} is a minimum of $\overline\beta_n$.  The right panel of
figure~\ref{fig:nmin} shows the contours in $\overline\beta_3$ for a 2-step
decay as a function of $m_1$ and $m_2$.

We can now formulate appropriate simplified models.  We see
that the direct decay of the LCP into the stau mediated by a
nearly mass-degenerate neutralino as well as the mass pattern
(\ref{eq:masspatt}) 
are reasonable 
benchmark mass patterns. We will display our results for three choices.
\begin{description}
\item[Model $\mathcal{A}$] The `direct decay'
via a nearly degenerate neutralino ($\mne\simeq\mstau$).
\item[Model $\mathcal{B}$]
The mass pattern (\ref{eq:masspatt}) for the $1$-step decay 
$\text{LCP}\to\neu\to\stau$, $\mne=\sqrt{\mlcp\mstau}$.
\item[Model $\mathcal{C}$] The mass pattern (\ref{eq:masspatt}) 
for the $3$-step decay $\text{LCP}\to\neu_2\to\s\ell\to\neu_1\to\stau$.
\end{description}
In all cases we will force the respective branching ratios to provide
the desired cascades (see section~\ref{sec:signal} for details).
We will assume symmetric decay chains, i.e., the
same cascade for both LCPs produced in an event.  We will briefly
discuss the issue of asymmetric chains in section~\ref{sec:sqresults}.

The model with pattern (\ref{eq:masspatt}) and the $2$-step decay
$\text{LCP}\to\neu_2\to\neu_1\to\stau$ turned out to lie completely
between $\mathcal{B}$ and $\mathcal{C}$ concerning the LHC 
sensitivity shown in section \ref{sec:sqresults}.  By this we implicitly 
checked also that the appearance of the heavy SM particle radiated
in the decay of the heavier neutralino does not change 
the qualitative picture.
Threshold effects are expected to be small due to the large
SUSY masses.\footnote{An exception could be a dominant 
decay chain with Higgsinos whose decay produces a
heavy Higgs. Here, threshold effects can be 
somewhat more important but are not expected to change the
picture significantly.
}

\section{Background estimation and selection criteria} \label{sec:bkgsel}

For the initial production of colored superparticles,
each event contains at least two jets, two staus and two
taus or tau neutrinos.  As the identification of taus it not very efficient, 
we do not include them in the signature. Furthermore, we will see that
the background rejection can already be saturated with staus
and jets alone.

In the detector, long-lived staus show up as muon-like particles, i.e.,
charged particles usually leaving the detector.  They can have a velocity
$\beta$ significantly below the speed of light, which allows one to
distinguish them from muons by virtue of a cut on $\beta$.  However, 
in some regions in parameter space many staus are produced with
a velocity close to~$1$.  These regions typically feature spectra with
large mass gaps $\mlcp-\mstau$.
Requiring hard jets can alleviate the drop in sensitivity
when $\beta$ approaches $1$, allowing
a relaxation of the cut on $\beta$. However, 
dropping the $\beta$ cut completely will 
lead to a
substantial loss of sensitivity due to a dramatic increase of the
(then unsuppressed) muon background.\footnote{%
A sufficient background rejection might be achievable without
a velocity discrimination if a very specific signature of SM radiation
is considered. However, such a search would introduce 
a strong model dependence, which is against the idea of this work.
}
Thus, we will always require identified staus.
If the staus stem from rather compressed spectra the jets are expected to 
be soft. On the other hand, the staus tend to be slow in these cases,
so identified staus alone suffice for a good sensitivity.

Since the SUSY particles are always produced in pairs, the largest
significance can be achieved by requiring \emph{two}
stau candidates in each event.
Therefore, we use the following signature for SUSY events:
\begin{itemize}
\item $2$ high-$\pt$, isolated muon-like particles passing a
velocity cut and
\item (optionally) $2$ high-$\pt$ jets.
\end{itemize}

\subsection{Background kinematics}\label{sec:background}

As background we consider all relevant SM processes providing 
two (isolated) muons. We will examine the behavior under several
kinematic cuts in this subsection. In the next subsection we will
discuss how the background is further reduced by a cut on the velocity.

We consider the Drell-Yan (DY) production of muons ($Z/\gamma\to\mu\mu$) 
and taus ($Z/\gamma\to\tau\tau$), di-boson production ($W^+W^-$, $WZ$ and $ZZ$),
$t\bar{t}$ production, single $t$ production ($tW$ plus $tb$) and
associated $Wb$ production, with jets from initial or final state radiation.
We calculated the cross section for DY production with
\textsc{FEWZ} \cite{Gavin:2010az} at next-to-NLO (NNLO) accuracy.
The cross sections for di-boson production \cite{Campbell:2011bn}, 
$t\bar{t}$ and single $t$ production \cite{Campbell:2004ch,Campbell:2005bb}
as well as associated $Wb$ production \cite{Campbell:2008hh,Caola:2011pz} 
were calculated via \textsc{MCFM} \cite{MCFMweb} at NLO precision.
We generated the events with \ME\ 5~\cite{Alwall:2011uj}. 
To regularize the collinear singularity and gain generator efficiency
we imposed the generator-level cuts $p_{\text{T}}^{\ell}>60\GEV$ 
(on both leptons) and $p_\text{T}^b>60\GEV$ (required for
at least one $b$ quark) 
in the normalization and event generation of the processes DY and $Wb$,
respectively.
The resulting cross sections are summarized in table \ref{tab:bkg}.

We performed showering and hadronization
with \PY~6~\cite{sjostrand-2006-0605}.
Since we will impose a selection criterion requiring two hard jets, the
distribution of the two leading jets should be reliable up to very 
high $\pt$.
Therefore, for the processes DY, $W^+W^-$, single $t$ and $Wb$
we include up to two additional jets in the matrix element simulation of \ME,
whereas for the processes $WZ$, $ZZ$\footnote{%
If both of the two hardest jets did not originate from the $W$ or $Z$, 
respectively, but from initial state radiation, this process would be just another 
correction to the DY process with an additional suppression by the weak 
coupling due to the production of the extra vector boson. Hence, we can easily
estimate that such a process cannot compete with the DY process, independent 
of the cuts we apply on the two jets and muons.
}
and $t\bar{t}$ we consider up to one additional jet---in the latter processes 
(at least) one of the two leading jets is expected to originate from the decay of 
a heavy SM particle. In the case of $Wb$,
one of the additional jets in the matrix element is allowed to be a $b$ jet
in order to include the $Wb\bar{b}$ contribution. 
In order to properly match the different contributions to the inclusive 
sample, which contains jets both from showering and from the matrix 
element, we applied the MLM matching procedure \cite{alwall-2008-53}
and chose $\texttt{xqcut}=p_{\text{T}}^{\text{jet,\,min}}\!=30\GEV$ 
and $\texttt{QCUT}=40\GEV$. We used the \textsc{Cteq6l1} PDF set \cite{Pumplin:2002vw}.

\begin{table}
\centering
\begin{tabular}{r   c  cccc} 
& & \multicolumn{4}{c}{$\sigma$ [pb]  } \vspace{0.2ex}\\
\cline{3-6} 
process & order &total  & selection\,1 & selection\,2 & selection\,3  \vspace{0.2ex} \\
\hline\vspace{-2ex}\\
DY$(\mu\mu)\!+\!X$ 		
& NNLO 	& 5.53	& 0.0038 		& 0.0971		& 0.0040  \\
DY$(\tau\tau)\!+\!X$ 	
& NNLO 	& 5.53	& $<\!0.0002$ 	&  $<\!0.0002$	& $<\!0.0002$	  \\
$W^+W^-\!+\!X$ 		
& NLO 	& 57.3 	& $<\!0.0002$ 	& 0.0015		& 0.0002  \\
$WZ\! +\!X$ 		
& NLO 	& 22.8 	& $<\!0.0002$ 	& 0.0005		&$<\!0.0002$  \\
$ZZ\!+\!X$ 		
& NLO 	& 7.92	& $<\!0.0002$  	& 0.0005 		&$<\!0.0002$  \\
$t\bar{t}\!+\!X$ 			
& NLO 	& 256	& 0.0019   	& 0.0034         	& $<\!0.0002$  \\
$t\!+\!X$		
& NLO 	& 96.5	& $<\!0.0002$  	& 0.0022 		&$<\!0.0002$  \\
$Wb\!+\!X$			
& NLO 	& 36.2	& $<\!0.0002$	&$<\!0.0002$	&$<\!0.0002$  \\
\hline\vspace{-2ex}\\
\multicolumn{3}{r}{$\Sigma\,| \,\text{before}\;\beta\text{-cuts}\,$} & 0.0069		& 0.1055		& 0.0054 \\
\multicolumn{3}{r}{$\Sigma\,| \,{\text{after}\;\beta\text{-cuts}}\,$}	& $4.8\times10^{-7}$	& $8.2\times10^{-8}$	& $<10^{-9} $
\end{tabular}
\caption{Cross sections for the considered background sources at the $8\TEV$
LHC\@. We display the total cross sections as computed
at the given precision (see text for details) as well as the cross
sections times efficiency obtained by applying the cuts selection criteria 1--3
defined in section \ref{sec:selection} but without cuts on the velocity
$\beta$.  The cuts on the velocity belonging to selection criteria 1--3
are only applied in the very last line.
}
\label{tab:bkg}
\end{table}

\begin{figure}
\centering
\setlength{\unitlength}{1\textwidth}
\begin{picture}(0.9,0.8)
  \put(0.0,0.4){
  \put(-0.025,0.025){\includegraphics[scale=1.11]{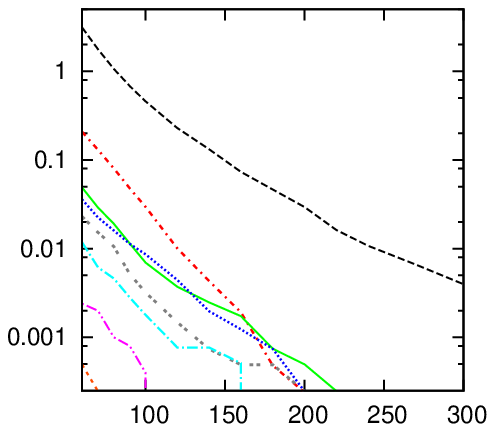}} 
  \put(0.248,0.0){\footnotesize $\pt^{\mu,\text{min}}$}
  \put(0.018,0.173){\rotatebox{90}{\footnotesize$\sigma\;[\pb\,]$}}
  }
 \put(0.43,0.4){
  \put(-0.025,0.025){\includegraphics[scale=1.11]{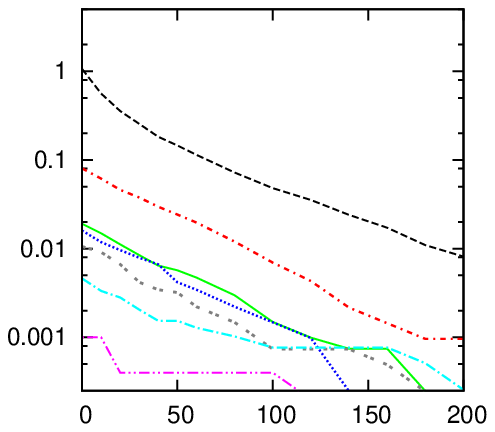}} 
  \put(0.248,0.0){\footnotesize $\Delta p_\text{T}^{\mu,\text{min}}$}
  \put(0.018,0.173){\rotatebox{90}{\footnotesize$\sigma\;[\pb\,]$}}
  }
 \put(0,0){
  \put(-0.025,0.025){\includegraphics[scale=1.11]{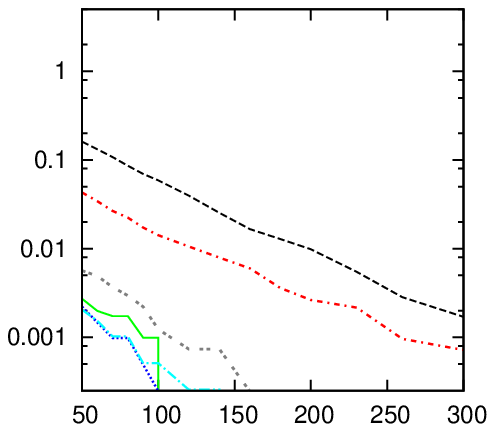}} 
  \put(0.248,0.0){\footnotesize $p_\text{T}^\text{jet,\,min}$}
  \put(0.018,0.173){\rotatebox{90}{\footnotesize$\sigma\;[\pb\,]$}}
  }
 \put(0.43,0){
  \put(-0.025,0.025){\includegraphics[scale=1.11]{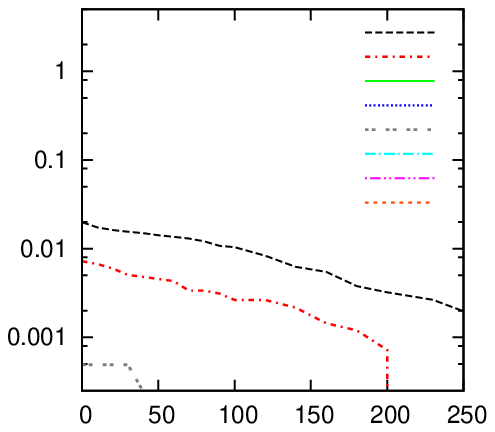}}
  \put(0.248,0.0){\footnotesize $\Delta p_\text{T}^\text{jet,\,min}$}
  \put(0.018,0.173){\rotatebox{90}{\footnotesize$\sigma\;[\pb\,]$}}
  \put(0.028,0.026){
    \put(0.1963,0.3076){\tiny DY$(\mu\mu)\!+\!X$}
  \put(0.2434,0.288971){\tiny  $t\bar{t}\!+\!X$}
  \put(0.25,0.270343){\tiny  $t\!+\!X$}
  \put(0.2193,0.251714){\tiny  $WW\!+\!X$}
  \put(0.225,0.233086){\tiny  $WZ\!+\!X$}
  \put(0.2308,0.214457){\tiny  $ZZ\!+\!X$}
    \put(0.1975,0.195829){\tiny  DY$(\tau\tau)\!+\!X$}
  \put(0.23,0.1772){\tiny  $Wb\!+\!X$}
  }
  }
\end{picture}
\caption{Inclusive cross section of the considered SM background processes 
as a function of various cuts at the $8\TEV$ LHC\@.
In all plots we imposed the isolation cut $\IREL<0.2$ and the pseudorapidity 
cut $|\eta|<2.4$ on each muon.
\emph{Top left:}~Requiring at least two muons with 
$p_\text{T}^\mu>p_\text{T}^{\mu,\text{min}}$ each. 
$p_\text{T}^{\mu,\text{min}}$ is varied.
\emph{Top right:}~Requiring at least two muons with $p_\text{T}^\mu>80\GEV$
and $\Delta p_\text{T}^\mu > \Delta p_\text{T}^{\mu,\text{min}}$. 
$\Delta p_\text{T}^{\mu,\text{min}}$ is varied.
\emph{Bottom left:}~Requiring at least two muons with 
$p_\text{T}^\mu>80\GEV$ and additionally at least two jets with
$p_\text{T}^\text{jet}>p_\text{T}^\text{jet,\,min}$ each. We vary 
$p_\text{T}^\text{jet,\,min}$.
\emph{Bottom right:}~Requiring at least two muons with 
$p_\text{T}^\mu>80\GEV$ and additionally at least two jets with
$p_\text{T}^\text{jet}>150\GEV$ each as well as
$\Delta p_\text{T}^\text{jet} > \Delta p_\text{T}^\text{jet,\,min}$.
We vary $\Delta p_\text{T}^\text{jet,\,min}$.
The key in the bottom right panel holds for all panels.
}
\label{fig:bkg}
\end{figure}

We passed the output of \PY\ to the detector simulation
\DP~1.9~\cite{Ovyn:2009tx} and applied a cut and count analysis 
on the lhco output of  \DP. The reconstruction
efficiency for each muon was set to 0.9. The  trigger
efficiency was conservatively set to $100\%$ for the background.
Figure \ref{fig:bkg} shows the cross sections for all considered background
processes as functions of various cuts on characteristic variables. These
variables are the transverse momentum of the muon $p_{\text{T}}^\mu$ and
of the jet $p_{\text{T}}^{\text{jet}}$, the difference of the $\pt$ of the two considered 
muons,
\beq
\Delta{p_{\text{T}}^\mu} = |p_{\text{T}}^{\mu,1}-p_{\text{T}}^{\mu,2}|\,,
\eeq
and of the two considered (hardest) jets,
\beq
\Delta{p_{\text{T}}^{\text{jet}}} = |p_{\text{T}}^{\text{jet},1}-p_{\text{T}}^{\text{jet},2}|\,.
\eeq
To reject the QCD background we require \emph{isolated} muons, i.e., $\IREL<\IREL^{\max}$,
where
\begin{equation} \label{eq:irel}
\IREL = 
\frac{\sum_{i\neq\mu}p_{\text{T},i}^\text{track}+\sum_{i\neq\mu}E_{\text{T},i}^\text{CAL}
}{p_{\text{T}}^\mu}
\end{equation}
is the relative isolation. In (\ref{eq:irel}) the sums are performed
over all objects within a cone of a given
$\Delta R \equiv \sqrt{\Delta\eta^2+\Delta\phi^2} = 0.3$ 
around the muon.\footnote{%
The energy deposition and track associated with this
particle itself are excluded from the sums.
Throughout this work we use the anti-$k_{\text{t}}$ jet clustering
algorithm \cite{Cacciari:2008gp}.
}

Except for the $Wb\!+\!X$ contribution the 
background is not sensitive to the precise choice of 
$\IREL^{\max}$ in a range from 0.02 to 1.
The same is true for the typical signal we will consider.
In contrast, $Wb+X$ drops significantly
with smaller $\IREL^{\max}$ and is practically 
irrelevant for the chosen value of $\IREL^{\max}=0.2$. 
Hard muons originating from $b\bar{b}$ are suppressed even more strongly
by the isolation cut and do not play any role here.
The same is expected for other QCD backgrounds.

The DY production is the
dominant background providing two hard isolated muons even with the 
requirement of two hard jets.
Without the requirement of hard jets, $t\bar{t}\!+\!X$ contributes a few
percent of the background, while with this requirement
its contribution constitutes up to approximately $15\%$.
The other sources are small and do not exceed a few percent 
in total.

\subsection{Discrimination via the velocity}\label{sec:velocity}

The main
difference between staus and muons is their velocity, which can be 
measured independently via the ionization loss in the tracker ($\D E/\D x$) 
and via a time-of-flight (ToF) measurement.
The relative uncertainty of the ToF measurement of muons (with 
$\beta\simeq1$) is approximately 0.048 in the ATLAS detector 
\cite{Aad:2011hz} and around 0.06 at CMS \cite{CMS-PAS-EXO-11-022}. 
The relative uncertainty of the velocity measurement via ionization loss is
smaller, around 0.035, but it is biased due to truncation effects 
\cite{CMS-PAS-EXO-10-004}. However, in general it will be possible to 
correct for this bias in an event-by-event analysis. The combination of the 
ToF and ionization loss measurement 
\cite{CMS-PAS-EXO-11-022,CMS-PAS-EXO-08-003}
yields an unambiguous and very robust measurement of the velocity.
Hence, we refer to this combination whenever possible. We estimated the 
relative uncertainty of the combined measurement by taking the weighted 
mean of the respective uncertainties for CMS, yielding 
$\sigma_\beta^{\text{rel}}\simeq0.032$.

A smearing of the velocity of muons is not included in the detector simulation
\DP. Assuming that there is no correlation between the velocity 
mismeasurement and the other observables\footnote{%
This is an obvious assumption for the background muons because a
deviation from $\beta=1$ does not originate from the physical process.
A dependence of the velocity measurement on the pseudorapidity
was reported to be small \cite{CMS-PAS-EXO-11-022}.
}
considered in the previous subsection, we treat the background 
rejection due to the velocity cut separately from the application of the 
kinematic cuts discussed there. In other words, for the background we 
first apply the cuts on $\IREL$, $\eta$, $p_{\text{T}}^{\text{jet}}$, $p_{\text{T}}^\mu$
$\Delta{p_{\text{T}}^\mu}$ and $\Delta{p_{\text{T}}^{\text{jet}}} $ to the generated
events and then multiply the resulting cross section by the background 
rejection factor $\R$ due to the velocity cut. To estimate $\R$ we assume 
a Gaussian smearing of the velocity with the respective width
$\sigma_\beta^\text{rel}$
given above.
In the case of the signal we refrain from smearing the velocity and use
the generator-level values (which were passed through \DP) to allow for
an event-based application of the cuts.

Since we always consider two stau candidates, we can combine the velocities
of the two stau candidates in different ways to formulate appropriate
cuts, which
lead to different factors $\R$. In the following, we denote the background 
rejection factor due to a cut on a \emph{single} muon by $\hat{r}_\beta$.
If the velocities of the two staus within one event tend to be correlated,
a cut on both stau candidates with the same $\beta_{\text{max}}$ will yield
the highest
sensitivity. We will denote this cut as
\beq
\beta^\Box< \beta_{\text{max}}^\Box\,.
\eeq
Since the mismeasurement of the velocity of two background muons
in one event is not correlated, this yields a background rejection factor 
of $\R= \hat{r}_\beta^2$.
If, in contrast, the velocity of the two staus in one event is strongly
uncorrelated, the cut
\beq
1- \sqrt{(1-\beta_1)^2 +(1-\beta_2)^2} \equiv \beta^\circ<\beta_{\text{max}}^\circ
\label{eq:betarad}
\eeq
yields a higher sensitivity on the signal. The background rejection factor is
$\R= \hat{r}_\beta$ in this case (for the considered Gaussian smearing).

\subsection{Lower limits on the velocity}\label{sec:lowbeta}

For staus with $\beta<0.6$ the efficiencies of the current triggers
at ATLAS and CMS drop significantly. In order to improve the trigger for
very slow staus the tracker
data has to be buffered in order to allow for a
recording of the tracker data 
in delay. Very slow staus are typically produced in scenarios with large 
$\mstau$. We therefore propose a recording of up to about four bunch 
crossings after a trigger by
muon-like particles with $\pt^\mu >300\GEV$ $(500\GEV)$ at the $8\TEV$ 
($14\TEV$) LHC run. We do not expect this to cause the recorded event rate 
to grow significantly. In section \ref{sec:selection} we will introduce
several selection criteria one of which assumes the proposed trigger
while the others simply require $\beta>0.6$.

Although staus suffer a high energy loss due to ionization of the detector 
material, they will often lose only a small fraction of their total energy.
Consequently, their velocity will stay approximately constant when
passing the detector. However, since the ionization loss increases with 
decreasing velocity the ionization loss will become relevant for the kinematics 
of the stau if the velocity falls below a critical value---staus might
then lose their kinetic energy 
completely and become trapped inside the detector. Since the traveling
range of charged particles in matter increases linearly with their mass, heavier
staus are more likely to pass the detector than lighter ones with the same velocity. 
On the other hand, very slow staus typically appear only if they are very heavy, so 
in conclusion stopped staus are rather the exception than the generic scenario.\footnote{%
We will examine the possibility of observing stopped staus at the LHC in 
section \ref{sec:stopped}.
}

However, to be able to record the tracks of the staus we have to make sure
that at least one stau reaches the muon chambers to fire the muon trigger.\footnote{%
This requirement can of course be relaxed if an event contains
enough SM particle radiation to fire the trigger. 
} 
Besides, both
have to pass the tracker to allow for an ionization loss measurement.
In order to account for this, we
used the approximate traveling range of a charged particle in the
detector material
given in \cite{0954-3899-37-7A-075021} (see 
section~\ref{sec:stopped} for details) and determined the minimal velocity that still 
ensures the required traveling range $R$ as a function of the mass of the stau, 
$\beta_{\min}^R(\mstau)$. The range $R$ was conservatively set to 
$1200\,\text{g}\,\text{cm}^{-2}$ and $12\,000\,\text{g}\,\text{cm}^{-2}$ for the requirement 
of passing the tracker and muon trigger, respectively. For a homogeneous detector 
material with density $\rho=8\,\text{g}\,\text{cm}^{-3}$, this corresponds to a 
path length of $1.5\,$m and 15\,m, respectively.

\subsection{Selection criteria} \label{sec:selection}

We will now introduce a set of three selection criteria,
each of which provides strong background rejection and high signal efficiency
in its domain in the considered parameter space. The selection criteria
are chosen in a complementary way such that the union of these
criteria will lead to high efficiencies throughout the whole parameter space
of the simplified models introduced in section \ref{sec:simplified}.

\begin{figure}[tbh]
\centering
\setlength{\unitlength}{1\textwidth}
\begin{picture}(0.82,0.54)
\put(0.0,0.01){
  \put(0.04,0.362){\includegraphics[scale=1.1]{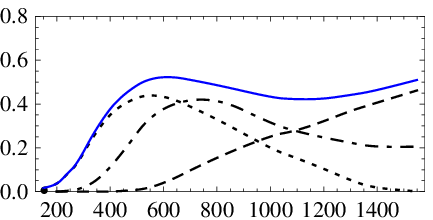}} 
  \put(0.04,0.206){\includegraphics[scale=1.1]{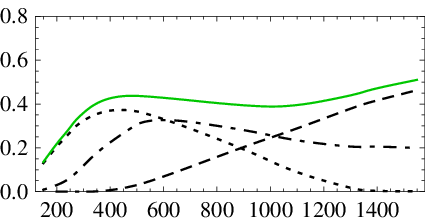}}   
  \put(0.04,0.034){\includegraphics[scale=1.1]{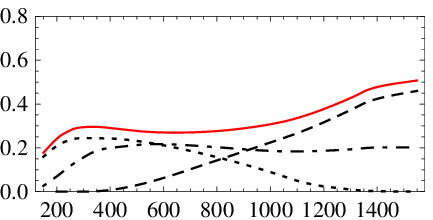}}   
  \put(0.167,0.0){\footnotesize $\mstau\;[\GEV\,]$}
  \put(0.001,0.392){\rotatebox{90}{\footnotesize efficiency}}  
  \put(0.0,0.234){\rotatebox{90}{\footnotesize efficiency}}
  \put(0.001,0.08){\rotatebox{90}{\footnotesize efficiency}}
  \put(0.235,0.377){\tiny sel.\,1}
  \put(0.314,0.388){\tiny sel.\,2}
  \put(0.314,0.42){\tiny sel.\,3}
  \put(0.08,0.474){\footnotesize $\mathcal{A}$}
  \put(0.08,0.318){\footnotesize $\mathcal{B}$}
  \put(0.08,0.162){\footnotesize $\mathcal{C}$}
  \put(0.25,0.474){\footnotesize $\,pp\,(8\TEV)$}
  }
\put(0.43,0.01){
  \put(0.04,0.362){\includegraphics[scale=1.1]{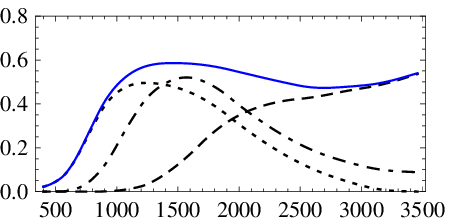}} 
  \put(0.04,0.206){\includegraphics[scale=1.1]{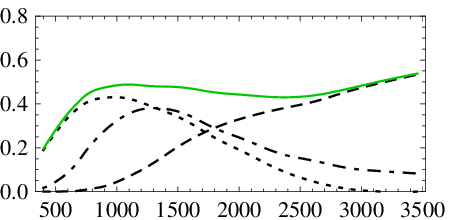}}   
  \put(0.04,0.034){\includegraphics[scale=1.1]{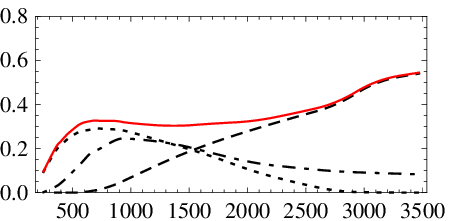}}   
  \put(0.167,0.0){\footnotesize $\mstau\;[\GEV\,]$}
  \put(0.001,0.392){\rotatebox{90}{\footnotesize efficiency}}  
  \put(0.0,0.234){\rotatebox{90}{\footnotesize efficiency}}
  \put(0.001,0.08){\rotatebox{90}{\footnotesize efficiency}}
  \put(0.23,0.376){\tiny sel.\,1}
  \put(0.314,0.393){\tiny sel.\,2}
  \put(0.314,0.43){\tiny sel.\,3}
  \put(0.24,0.474){\footnotesize $\,pp\,(14\TEV)$}
  }
\end{picture}
\caption{Efficiencies of the selection criteria 1 to 3 as functions
of $\mstau$ for fixed squark and gluino masses $\mgo=\msq=
1600\GEV\;(3600\GEV)$ for the $8\TEV$ ($14\TEV$) LHC and for 
the simplified models~$\mathcal{A}$~(\emph{top panels}), $\mathcal{B}$ 
(\emph{middle}) and $\mathcal{C}$ (\emph{bottom}). Trigger and
reconstruction efficiencies are included. The solid curve denotes
the union of all three cuts,
i.e., the
efficiency resulting from selecting all events satisfying at least one of
the selection criteria.
}
\label{fig:efficiencies}
\end{figure}

In the following, each stau candidate is understood to pass the
isolation criterion $\IREL<0.2$ and to lie within a pseudorapidity range $|\eta|<2.4$. 
When we cut on the transverse momentum $\pt^{2i}$, we require \emph{two} 
particles $i$ that both pass the cut. The values given 
below are valid for the $8\TEV$ ($14\TEV$) LHC analysis.
The selection criteria are:

\begin{enumerate}
\item Two stau candidates passing the muon chambers,
\beq\label{eq:sel1a}
\pt^{2\mu} >80\GEV\;\,(240\GEV)\, ,\quad \Delta{p_{\text{T}}^\mu}>50\GEV\;\,(70\GEV)\, ,\quad\beta^\circ< 0.86\,,
\eeq
and two jets,
\beq \label{eq:sel1b}
\pt^{2\text{jet}}>200\GEV\;\,(400\GEV)\,.
\eeq
One stau candidate is required to
fire the muon trigger `in time',
\beq
\beta
>0.6\,. 
\eeq

\item Two stau candidates passing the muon chambers,
\beq\label{eq:sel2a}
\pt^{2\mu} >150\GEV\;\,(360\GEV)\, ,\quad\beta^\Box< 0.88\,,
\eeq
and 
\beq \label{eq:sel2b}
\beta
>0.6\,
\eeq
for one stau.

\item Two stau candidates passing the tracker,
\beq\label{eq:sel3a}
\pt^{2\mu} >300\GEV\;\,(500\GEV)\, ,\quad\beta^\Box< 0.73\,,
\eeq
one of which has to pass the
muon trigger chambers,
\beq \label{eq:sel3b}
\beta_1>\beta_{\min}^{\text{tracker}}(\mstau)\, ,\quad\beta_2>\beta_{\min}^{\text{trigger}}(\mstau)\,.
\eeq
This selection criterion assumes the modified trigger setup proposed in section
\ref{sec:lowbeta}.
\end{enumerate}
Figure \ref{fig:efficiencies} shows the efficiencies of the selection
criteria 1 to 3 for exemplary mass slices of the simplified models
$\mathcal{A}$, $\mathcal{B}$ and $\mathcal{C}$. 
For all simplified models, selection criterion 1 is most efficient if
the stau is sufficiently light.  The large mass difference between LCP
and stau ensures the production of high-$\pt$ jets that enable a very
good background rejection already in combination with the relatively
loose velocity cut on $\beta^\circ$, thus cutting away a smaller part of
the signal than with the cut on $\beta^\Box$.  For heavier staus,
high-$\pt$ jets are no longer guaranteed.  Consequently, selection
criterion 2 becomes more efficient, relying on the cut on $\beta^\Box$ to
discriminate against muons with a mismeasured velocity.  If the stau is
very heavy, many events will contain slow staus that do not pass the cut
$\beta > 0.6$.  However, due to their large mass they have a very large
$\pt$.  Hence, selection criterion 3 is optimal for this part of the
parameter space.

The background rejection obtained with these cuts is summarized in
table~\ref{tab:bkg} for the $8\TEV$ LHC\@. The relative importance of
the background sources is similar for the $14\TEV$ run. 
The $14\TEV$ cuts provide a stronger background suppression
as required by the larger integrated luminosity considered.

\section{Exploring the parameter space}\label{sec:expl}

\subsection{Computation of the signal}\label{sec:signal}

To estimate the projected LHC reach for the simplified models described in 
section \ref{sec:simplified} we performed a full-fledged Monte Carlo study.
Here we briefly sketch the computational steps of the analysis. The analysis 
was performed for the $8\TEV$ and $14\TEV$ LHC run.
For the sake of saving computing time we factorized production and decay. 

For the production we built up a grid of generator-level event files in the 
$\mgo$-$\msq$ plane. We considered the production channels $\go\go$, 
$\go\sq$, $\sq\sq$ and $\sq\overline\sq$ with a common squark mass for 
$\sq=\s u,\s d,\s s,\s c,\s b$.
We computed the production cross sections for the different channels 
at NLO precision via \textsc{Prospino~2}
and simulated the events with the tree-level generator \ME~5. 
Since the size of NLO corrections differs significantly between the
considered production channels,
we performed a channel-wise normalization, i.e., we treated
each production channel separately throughout the analysis chain.
We generated a total of $30\,000$ events per mass point, apportioned
between the four production channels according to their fraction of the
total cross section. We used the \textsc{Cteq6l1} PDF set. 

In a second step we passed the \ME\ events to \PY~6 to perform
the decay of the SUSY particles (and the showering and hadronization
of SM particles).
For each point on the $\mgo$-$\msq$ grid, this allows for a variation
of the spectrum below the LCP\@. We computed the decay widths and 
branching ratios (BR) via \textsc{SDECAY} \cite{Muhlleitner:2003vg}.
The minimal decay chain $\text{LCP}\to\neu\to\stau$ was obtained
by decoupling all other SUSY particles. The decay chain 
$\text{LCP}\to\neu_2\to\s\ell\to\neu_1\to\stau_1$ was
enforced by computing the BR down to the $\neu_2$ via
\textsc{SDECAY} and adjusting the following BR accordingly.

The use of \PY~6 for the decay of SUSY particles implies certain
approximations whose validity we have to justify.
\PY~6 factorizes the cascade into decays of on-shell
particles, using the narrow width approximation (i.e., each decay width
is much smaller than the corresponding mass difference),
and neglects spin correlations of fermions in the chain.
Since we are interested in a systematic scan of the free parameters 
of the simplified models
including regions where masses of sparticles in the decay chain
are nearly mass degenerate, we have to ensure the validity of 
the use of factorization here. 

The only kinematical cuts this analysis strongly relies on concern the $\beta$ 
of the staus, the $\pt$ of the staus and the $\pt$ of the two hardest jets 
and combinations thereof.
Consider the case where two SUSY particles in the cascade are very close in
mass, i.e., their masses are much larger than the mass difference and much larger
than the mass of the radiated SM particle. 
In this case the daughter sparticle
basically inherits the kinematics of the mother sparticle and the SM particle 
appears as soft radiation. Its distribution might not be
described well by the simulation. However, 
we do not cut on SM particles other than the jets. As the jets have to survive very 
severe $\pt$ cuts, jets from such degenerate decays are very unlikely
to contribute to our signal. 
We have explicitly checked the $\pt$ distributions of the two hardest jets and
the $\pt$ and $\beta$ distributions of the staus against a full matrix element 
simulation by \WZ\ \cite{Kilian:2007gr}
and found reasonably good agreement for different setups in which 
we considered the decay $\sq \to q \s\chi^0 \to q \tau \stau_1$ for (nearly) degenerate
$\msq$ and $\mne$, as well as nearly degenerate $\mne$ and $\mstau$.
The results differ by at most $10\%$. The resulting error on the 
efficiencies of the selection criteria which enters our final results is estimated 
to be even smaller, and is especially small compared to the
dominant uncertainties, which are PDF and scale uncertainties that 
are introduced via the production cross section (for an error estimation see
section \ref{sec:error}). Thus, factorization is well-justified when relying 
on the considered observables.\footnote{%
Relative angular distributions are affected more strongly
by the applied approximation. Thus, larger deviations are present in
$\Delta R$.  However, the dependence of our results on
$\IREL$ is very weak, as we stated in section \ref{sec:background}. 
Consequently, the results are not affected noticeably.
}

We included up to one additional jet in the matrix element. This introduces
a source of potential double counting. Gluino-gluino production with 
an additional jet contains, for instance, a diagram with an intermediate squark. If the
squark is on-shell, this contribution is equivalent to the lowest order gluino-squark
production followed by the decay of the squark. To account for this, \textsc{Prospino}
removes contributions from on-shell intermediate squarks and gluinos. Accordingly, we removed
the same diagrams in the event generation in the course of the matching
procedure in \PY. 
As for the background, we applied the
MLM matching procedure in order to match properly the jets
from the matrix element and showering.

We passed the output of \PY\ to the detector simulation
\DP~1.9. To account for long-lived staus we applied minor
modifications on \DP. The reconstruction
efficiency for each stau was set to $0.9$. We assumed a trigger
efficiency of $90\%$ \cite{KoljasThesis}.

\subsection{LHC reach for a common squark mass}\label{sec:sqresults}

We estimate the LHC's sensitivity to observe or
exclude the introduced simplified models as a function of the free parameters $\mgo$, 
$\msq$ and $\mstau$. We consider a common squark mass $\msq$ in this 
section. As discussed in section \ref{sec:prod} the production is dominated by the 
first-generation squarks. Consequently, the derived limits can be interpreted as 
limits on the masses of these squarks. In section \ref{sec:lightstop} we will consider 
the case of a light stop dominating the production cross section.

As shown in figure \ref{fig:efficiencies} the efficiencies of the selection 
criteria are typically about $0.5$. On the other hand, as shown in table \ref{tab:bkg},
the background expectation is reduced by these cuts to less than $10^{-2}$ events.
Accordingly, a $95\%$ C.L.\ exclusion\footnote{%
We apply the CL$_{\text{S}}$ method here.}
can always be claimed if no events are observed
while three are expected.
That is, the background rejection is saturated with these cuts.
In such a situation the analysis is not very sensitive to the precise background 
cross section anymore (see e.g.\ \cite{Heisig:2011dr}).
Furthermore, discovery can also be claimed on the basis of
very few events and the results for the exclusion curves represent an
(typically even conservative) estimation of the $5\sigma$ discovery potential.
These are typical features of the search for heavy stable charged sparticles.

\begin{figure}[btp]
\centering
\setlength{\unitlength}{1\textwidth}
\begin{picture}(0.85,0.81)
\put(0.0,0.4){
  \put(0.04,0.035){\includegraphics[scale=1.1]{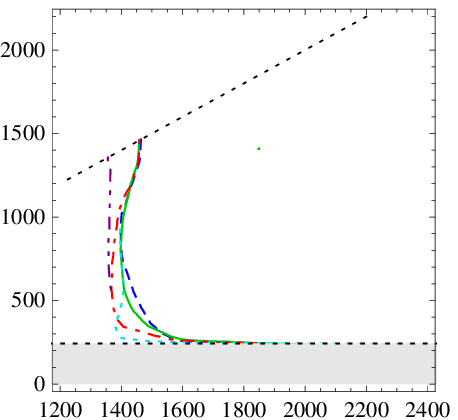}}
  \put(0.14,0.0){\footnotesize $\mgo=\tfrac 1 2 \msq\; [ \GEV \, ] $}
  \put(0.0,0.14){\rotatebox{90}{\footnotesize $\mstau\;[\GEV\,]$}}
  \put(0.09,0.316){\footnotesize $\,pp\,(8\TEV)$}
  \put(0.09,0.286){\footnotesize $\int\!\mathcal{L}=16\ifb$}
  \put(0.136,0.073){\tiny minimal DY production limit}
  }
 \put(0.43,0.4){
  \put(0.04,0.035){\includegraphics[scale=1.1]{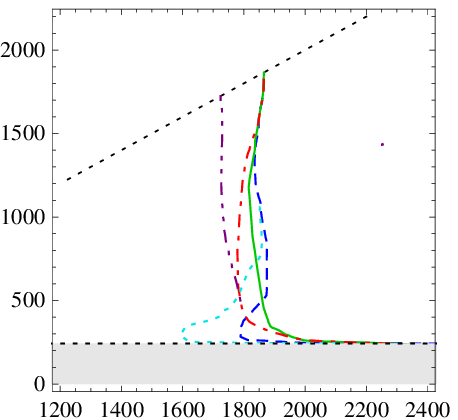}}   
  \put(0.15,0.0){\footnotesize $\mgo=\msq\; [ \GEV \, ] $}
  \put(0.0,0.14){\rotatebox{90}{\footnotesize $\mstau\;[\GEV\,]$}}
  \put(0.09,0.316){\footnotesize $\,pp\,(8\TEV)$}
  \put(0.09,0.286){\footnotesize $\int\!\mathcal{L}=16\ifb$}
  \put(0.136,0.073){\tiny minimal DY production limit}
  }
\put(0.0,0.0){
  \put(0.04,0.035){\includegraphics[scale=1.1]{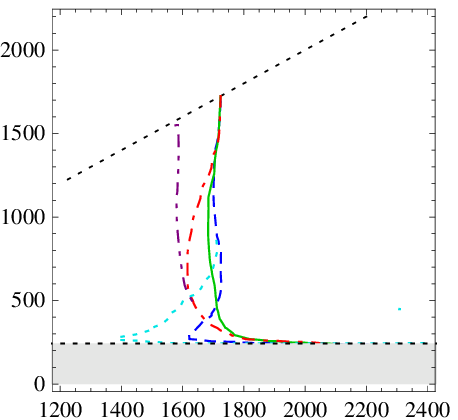}}
  \put(0.14,0.0){\footnotesize $\msq=\tfrac 1 2 \mgo \; [ \GEV \, ] $}
  \put(0.0,0.14){\rotatebox{90}{\footnotesize $\mstau\;[\GEV\,]$}}
  \put(0.09,0.316){\footnotesize $\,pp\,(8\TEV)$}
  \put(0.09,0.286){\footnotesize $\int\!\mathcal{L}=16\ifb$}
  \put(0.136,0.073){\tiny minimal DY production limit}
  }
\put(0.43,0){
  \put(0.04,0.035){\includegraphics[scale=1.1]{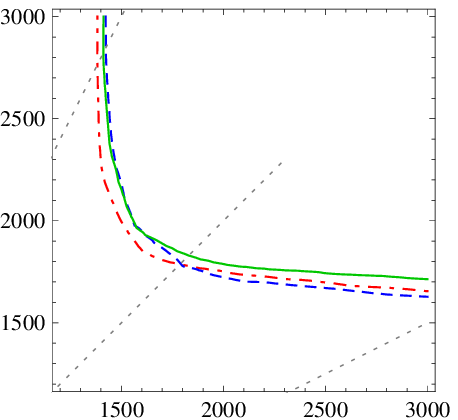}}   
  \put(0.18,0.0){\footnotesize $\mgo\; [ \GEV \, ] $}
  \put(0.0,0.14){\rotatebox{90}{\footnotesize $\msq\;[\GEV\,]$}}
  \put(0.215,0.316){\footnotesize $\,pp\,(8\TEV)$}
  \put(0.215,0.286){\footnotesize $\int\!\mathcal{L}=16\ifb$}
  \put(0.215,0.26){\tiny along minima in}
  \put(0.215,0.244){\tiny $\mstau$-variation}
  }
\end{picture}
\caption{Projected LHC sensitivity ($95\%\,\textnormal{CL}_\textnormal{s}$ 
exclusion and approximate $5\sigma$ discovery reach, see text)  for the 
models $\mathcal{A}$ (blue dashed), 
$\mathcal{B}$ (green solid) and $\mathcal{C}$ (red dot-dashed), as well as 
$\mathcal{A}$ (cyan dotted) and $\mathcal{C}$ (purple dot-dot-dashed) for
a reduced set of selection criteria (see text for details). A common squark
mass $\msq$ is assumed. In the lower right panel 
the curves represent the minima in the sensitivity with respect to the variation
of $\mstau$. 
}
\label{fig:sensitivity8}
\end{figure}

\begin{figure}
\centering
\setlength{\unitlength}{1\textwidth}
\begin{picture}(0.85,0.4)
\put(0.0,0.0){
  \put(0.04,0.035){\includegraphics[scale=1.1]{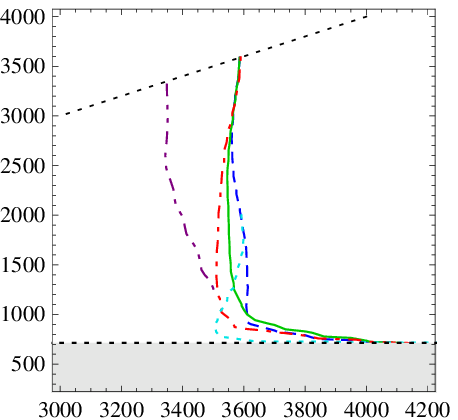}}
  \put(0.15,0.0){\footnotesize $\msq=\mgo \; [ \GEV \, ] $}
  \put(0.0,0.14){\rotatebox{90}{\footnotesize $\mstau\;[\GEV\,]$}}
  \put(0.23,0.286){\footnotesize $\,pp\,(14\TEV)$ }
  \put(0.23,0.256){\footnotesize $\int\!\mathcal{L}=300\ifb$}
  \put(0.136,0.073){\tiny minimal DY production limit}
  }
 \put(0.43,0){
  \put(0.04,0.035){\includegraphics[scale=1.1]{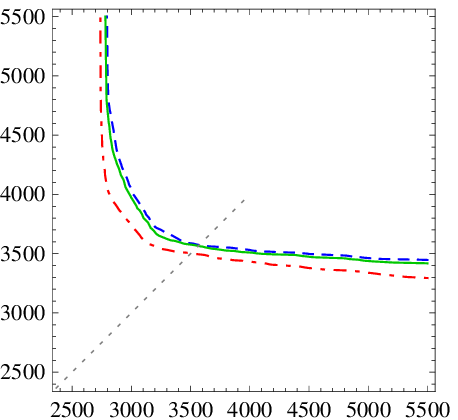}}   
  \put(0.18,0.0){\footnotesize $\mgo\; [ \GEV \, ] $}
  \put(0.0,0.14){\rotatebox{90}{\footnotesize $\msq\;[\GEV\,]$}}
  \put(0.215,0.316){\footnotesize $\,pp\,(14\TEV)$}
  \put(0.2145,0.286){\footnotesize $\int\!\mathcal{L}=300\ifb$}
 \put(0.215,0.26){\tiny along minima in}
 \put(0.215,0.244){\tiny $\mstau$-variation}
  }
\end{picture}
\caption{Projected LHC sensitivity ($95\%\,\textnormal{CL}_\textnormal{s}$ 
exclusion and approximate $5\sigma$ discovery reach, see text)  for the 
models $\mathcal{A}$ (blue dashed), 
$\mathcal{B}$ (green solid) and $\mathcal{C}$ (red dot-dashed), as well as 
$\mathcal{A}$ (cyan dotted) and $\mathcal{C}$ (purple dot-dot-dashed) for
a reduced set of selection criteria (see text for details). A common squark
mass $\msq$ is assumed. In the right panel the curves represent the minima 
in the sensitivity with respect to the variation of $\mstau$. 
}
\label{fig:sensitivity14}
\end{figure}

Figures \ref{fig:sensitivity8} and \ref{fig:sensitivity14} show the resulting 
sensitivity for the $8\TEV$ and $14\TEV$ LHC run, respectively, 
for the three simplified models introduced in section \ref{sec:decay}.
We visualize the variation of $\mgo$, $\msq$ and $\mstau$ by showing slices
of the parameter space.
In the plots showing $\mgo$-$\mstau$ and $\msq$-$\mstau$ planes, a
fixed ratio $\msq/\mgo$ is assumed.  In the plots of the
$\mgo$-$\msq$ plane,
we draw the sensitivity curves by conservatively 
choosing the stau mass that yields the smallest sensitivity at each 
point of the plane.
In addition to strong production and decay,
we include the production of staus by the direct DY process.
In order to derive conservative limits we considered the stau mixing
angle that yields the smallest cross section \cite{Heisig:2011dr}.\footnote{%
A significantly larger contribution of directly produced staus is possible, see e.g.\
\cite{Lindert:2011td}.}
This contribution is always present and depends on the stau mass only.\footnote{%
Similarly one could think of including the direct production of the intermediate sparticles
that appear in the cascades. However, since there is no Drell-Yan production for pairs
of pure binos the cross section can always be rendered negligible by an appropriate
choice of the neutralino mixing. Hence, at least in the minimal decay chain, there is no
such conservative minimal contribution. Including a minimal direct production cross 
section of the intermediate sparticles in longer decay chains would have a certain 
effect on the results in some of the considered regions in parameter space. 
However, this would run counter to the idea of this work.}

In model $\mathcal{A}$ (blue dashed lines) the decay $\text{LCP}\to\neu$
leads to hard jets and potentially highly boosted staus.
For moderate mass gaps $\mlcp-\mstau$ the staus are well-distinguishable
from muons. The additional jet signature leads to a slight
enhancement of the significance. For larger $\mlcp-\mstau$
a large number of staus are rejected by the velocity cut and the significance 
drops sharply, despite the fact that the jets become harder. 
This effect would hide the scenario very effectively
from our selection criteria if it
were not for the direct production, which increases the sensitivity for
lower $\mstau$
(see figure~\ref{fig:efficiencies}, where we did not include the direct production).
Thus, DY production allows us to cover the parameter
space with a mostly model-independent search. Without
it, we were forced to introduce dedicated searches for each
occurring topology to be able to cover the small $\mstau$ region.
For stau masses just above the region of dominant DY production 
(and for $\mgo\gtrsim \msq$) the sensitivity for model $\mathcal{A}$ reaches
a minimum (see upper right and lower left panel of figure
\ref{fig:sensitivity8}).
Higher luminosities push the corresponding mass reach up and cause the 
production to be closer to threshold.
As a consequence, the minimum disappears
in the projected sensitivity for the $14\TEV$ $300\ifb$ run (see left
panel of figure~\ref{fig:sensitivity14}).
The cyan dotted curves show the sensitivity
after dropping selection criterion 1, i.e., without taking the jet
signature into account. The diminished sensitivity
illustrates the importance of the additional 
jet signature in this region.

Model $\mathcal{C}$ has a sensitivity minimum in the intermediate 
range of $\mlcp-\mstau$. Many staus are rejected by the 
upper velocity cut in this region. Furthermore, compared to
model $\mathcal{A}$ fewer hard jets are produced that could
compensate for this effect.

As expected from the discussion in section \ref{sec:decay},
either $\mathcal{A}$ or $\mathcal{C}$ yields the minimal
sensitivity for all values of $\mgo$, $\msq$ and $\mstau$.
Model $\mathcal{B}$ behaves more moderately,
resulting in a sensitivity in between those of the other models
at most points.

For small mass gaps $\mlcp-\mstau\to0$ all three models
give the same efficiency. The mass pattern of the intermediate
sparticles does not play a role in this regime.
The staus basically inherit the velocity and angular distribution
of the produced colored sparticles.
In this parameter region lower limits on the 
velocity become important. As stated in section
\ref{sec:lowbeta} current trigger restrictions cause the loss of
events in which both staus have a velocity $\beta\lesssim0.6$.
For small mass gaps this loss is significant.
This is illustrated by the  purple dot-dot-dashed
curves that show the sensitivity for model $\mathcal{C}$ after dropping
selection criterion 3, which assumes buffering of the tracker 
data in order to be able to record several bunch crossings in delay.
Especially for the $14\TEV$ run and $300\ifb$ luminosity, the implementation
of such a trigger enhances the sensitivity significantly
(see left panel of figure \ref{fig:sensitivity14}).

In all the plots, the LHC sensitivities for the simplified models $\mathcal{A}$, 
$\mathcal{B}$ and $\mathcal{C}$ span a relatively narrow band although the 
mass patterns of the models are radically different. Furthermore, the overall 
dependence of the sensitivity on $\mstau$ is moderate. 
This situation is very different from the one in a missing $E_{\text{T}}$ search 
like in neutralino LSP scenarios. There the sensitivity depends much more on 
the intermediate spectrum and on the mass of the LSP and it is more difficult 
to cover the limiting cases with appropriate simplified models.
This shows that the simplified model approach is very suitable for the long-lived
stau scenario.
In the $\mgo$-$\msq$ plane, where we took the most conservative choice
for $\mstau$ at each point, the simplified models $\mathcal{A}$--$\mathcal{C}$ 
lie even more closely together. From the $\mgo$-$\msq$ plane plots
we can derive conservative projected limits on the gluino and squark masses 
in the common mass scenario.
With $\int\!\mathcal{L}=16\ifb$ at $8\TEV$ we expect gluino and squark masses 
of $\mgo\lesssim 1.4\TEV$ and $\msq\lesssim1.6\TEV$ to be either excluded
or discovered.
With $\int\!\mathcal{L}=300\ifb$ at $14\TEV$ we are sensitive to
$\mgo\lesssim 2.6\TEV$ and $\msq\lesssim3.3\TEV$.
These limits allow for a completely model-independent interpretation
with respect to all SUSY parameters that are not specified in this setup.

As already mentioned in section \ref{sec:decay},
asymmetric decay chains, i.e., one short and one long decay chain
in one event, will also appear in realistic models.
The selection criteria we imposed are not dedicated to such chains
and asymmetric cascades will partly fail to be selected by these
criteria. However, we expect that a dedicated extension of the
selection criteria will provide equally high signal-to-background ratios.
One could for instance require one very hard jet and an even 
stronger asymmetry in the stau kinematics. 

Another aspect we noted in section \ref{sec:decay} is the
presence of heavy SM particle radiation in the 2-step 
decay LCP$\to\neu_2\to\neu_1\to\stau$. To check whether
threshold effects might affect the sensitivity significantly, we
computed the corresponding curves for this 2-step decay. 
We explicitly allowed for the $3$-body decay of the heavier neutralino 
into the lightest neutralino, which occurs once $m_{\neu_2}-m_{\neu_1}<m_Z$.
The sensitivity curves for the 2-step decay lie completely between the 
3-step decay (model $\mathcal{C}$) and the 1-step decay (model $\mathcal{B}$)
in all plots of figures \ref{fig:sensitivity8} and \ref{fig:sensitivity14}.

As another check of the generality of the introduced simplified models,
we considered an inverted mass hierarchy in the simplified model
$\mathcal{C}$, $m_{\s\ell}<\mne$.  In this case a 3-body slepton decay
occurs.
In order to perform these computations we
used an extension of the \textsc{SDECAY} package \cite{Kraml:2007sx} 
allowing for the 3-body decays of sleptons. We found that for these spectra
the picture drawn here does not change.

\subsection{LHC reach for a light stop} \label{sec:lightstop}

As a complementary limit we will now consider the case of a light stop 
$\sTop_1$ and all other squarks decoupled. We set the gluino mass to 
$\mgo=3\mstop$. In this setup processes other than 
stop-antistop production are negligible.
Although the signature of the decaying stops might potentially
provide a larger significance with a dedicated selection criterion,
the benefit is expected to be marginal due to the high efficiencies
that are already provided by the introduced cuts. Therefore,
we refrain from introducing a dedicated selection here.

\begin{figure}
\centering
\setlength{\unitlength}{1\textwidth}
\begin{picture}(0.85,0.4)
\put(0.0,0.0){
  \put(0.04,0.035){\includegraphics[scale=1.1]{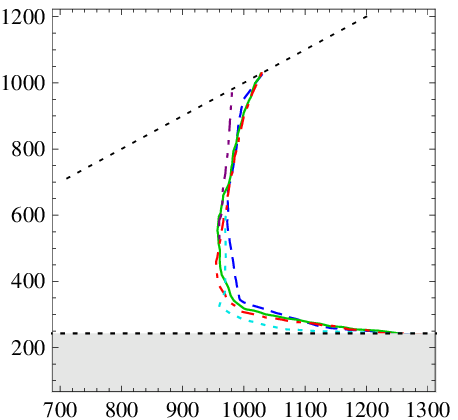}}
  \put(0.17,0.0){\footnotesize $\mstop\; [ \GEV \, ] $}
  \put(0.0,0.14){\rotatebox{90}{\footnotesize $\mstau\;[\GEV\,]$}}
  \put(0.09,0.32){\footnotesize $\,pp\,(8\TEV)$}
  \put(0.09,0.29){\footnotesize $\int\!\mathcal{L}=16\ifb$}
  \put(0.136,0.073){\tiny minimal DY production limit}
  }
 \put(0.43,0){
  \put(0.04,0.035){\includegraphics[scale=1.1]{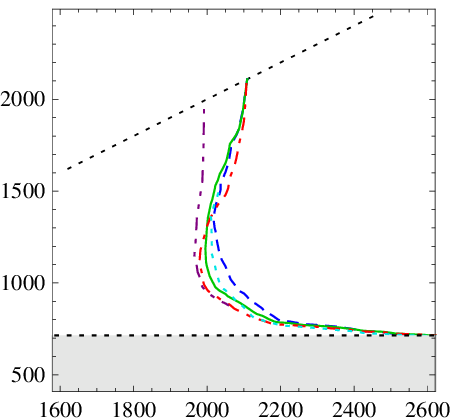}}   
  \put(0.17,0.0){\footnotesize $\mstop\; [ \GEV \, ] $}
  \put(0.0,0.14){\rotatebox{90}{\footnotesize $\mstau\;[\GEV\,]$}}
  \put(0.09,0.32){\footnotesize $\,pp\,(14\TEV)$}
  \put(0.09,0.29){\footnotesize $\int\!\mathcal{L}=300\ifb$}
  \put(0.136,0.073){\tiny minimal DY production limit}
  }
\end{picture}
\caption{Projected LHC sensitivity ($95\%\,\textnormal{CL}_\textnormal{s}$ 
exclusion and approximate $5\sigma$ discovery reach,
see section~\ref{sec:sqresults}) 
for a light stop $\sTop_1$ and decoupled
squarks and gluinos otherwise. 
As in figures \ref{fig:sensitivity8} and \ref{fig:sensitivity14}
the lines denote the models $\mathcal{A}$ (blue dashed), 
$\mathcal{B}$ (green solid) and $\mathcal{C}$ (red dot-dashed), as well as 
$\mathcal{A}$ (cyan dotted) and $\mathcal{C}$ (purple dot-dot-dashed) for
a reduced set of selection criteria (see section~\ref{sec:sqresults} for details).
}
\label{fig:lightstop}
\end{figure}
Figure \ref{fig:lightstop} shows the sensitivity in the 
$\mstop$-$\mstau$ plane. According to the lower production cross
section for stops, the stop masses that are in reach of the LHC are
smaller than the accessible squark masses that we have found in the common 
squark mass scenario considered in the previous section. 
The stau mass that is accessible via direct stau production
is the same, of course.
Correspondingly, the gap between the LCP mass
and the lowest stau masses for which the production via cascades is
dominant is considerably smaller. Therefore, the effect of the intermediate
spectrum becomes less important and the curves for models 
$\mathcal{A}$--$\mathcal{C}$
lie even more closely together than in the previous section.
In particular, the minimum of the sensitivity of model $\mathcal{A}$
just above the region of dominant direct DY production has disappeared.
The velocity distribution of the staus varies less significantly with the stau 
mass in this scenario.

For $\mstop>\mgo$ we find the same sensitivity to $\mgo$ as in the 
large-$\msq$ limit in the previous section.
In conclusion,
with $\int\!\mathcal{L}=16\ifb$ at $8\TEV$ we expect gluino and stop masses of 
$\mgo\lesssim 1.4\TEV$ and $\mstop\lesssim950\GEV$ to be either excluded
or discovered.
With $\int\!\mathcal{L}=300\ifb$ at $14\TEV$ we are sensitive to
$\mgo\lesssim 2.6\TEV$ and $\mstop\lesssim2\TEV$.

\subsection{Uncertainties} \label{sec:error}

We shall briefly discuss the theoretical uncertainties
of the cross section computations here. 
We expect these uncertainties to be the most important ones.
Further uncertainties arise from the event generation and from
the simplified detector simulation.
We only consider 
the error implied by the scale dependence of the production cross section.
\begin{figure}[tbhp]
\centering
\setlength{\unitlength}{1\textwidth}
\begin{picture}(0.85,0.4)
\put(0.0,0.0){
  \put(0.04,0.035){\includegraphics[scale=1.1]{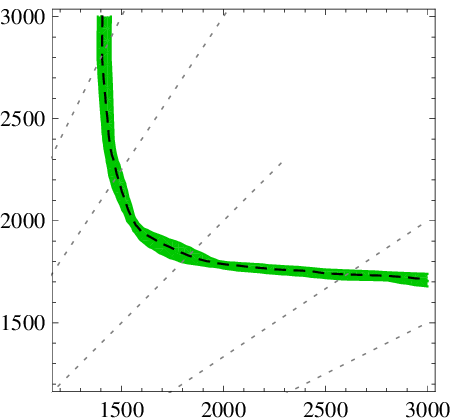}}
  \put(0.18,0.0){\footnotesize $\mgo\; [ \GEV \, ] $}
  \put(0.0,0.14){\rotatebox{90}{\footnotesize $\msq\;[\GEV\,]$}}
  \put(0.22,0.316){\footnotesize $\,pp\,(8\TEV)$}
  \put(0.22,0.286){\footnotesize $\int\!\mathcal{L}=16\ifb$}
 \put(0.22,0.256){\footnotesize scale var.}
  }
 \put(0.43,0){
  \put(0.04,0.194){\includegraphics[scale=1.1]{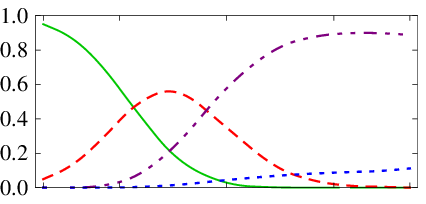}}   
  \put(0.04,0.034){\includegraphics[scale=1.1]{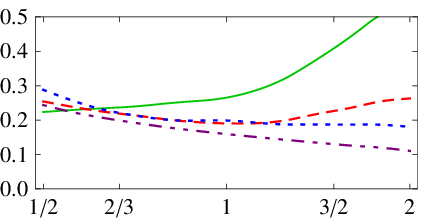}}   
  \put(0.18,0.0){\footnotesize $\mgo/\msq $}
  \put(0.0,0.248){\rotatebox{90}{\footnotesize $\sigma_i/\Sigma\sigma_i$}}
  \put(0.001,0.082){\rotatebox{90}{\footnotesize rel. error}}
  \put(0.32,0.313){\tiny $\sq\sq$}
  \put(0.32,0.24){\tiny $\sq\overline\sq$}
  \put(0.16,0.3){\tiny $\go\sq$}
  \put(0.08,0.305){\tiny $\go\go$}
  }
\end{picture}
\caption{\emph{Left:}~LHC sensitivity in the 
$\mgo$-$\msq$ plane for model $\mathcal{B}$ (black dashed curve, same as 
the green solid curve shown in the lower right panel of figure \ref{fig:sensitivity8}) 
and its error band according to the renormalization and factorization scale variation 
(green shaded region).
\emph{Upper right:}~Fractional contributions of the considered 
production channels to the total production cross section of
colored sparticles, parametrized by the ratio $\mgo/\msq$, which is varied
along the sensitivity limit shown in the left panel.
\emph{Lower right:}~Relative errors in the cross sections
due to scale variation for the considered production channels, again
changing $\mgo/\msq$ along the sensitivity limit.
}
\label{fig:scale}
\end{figure}
The error introduced by PDF uncertainties is roughly of the same order. 
Uncertainties in the strong coupling $\alpha_{\text{s}}$
are expected to be somewhat smaller. A detailed discussion of the
errors relevant for the production of colored sparticles at the LHC can be found in 
\cite{Beenakker:2011fu}.

In the left panel of figure \ref{fig:scale} we exemplarily show again the 
$8\TEV$ LHC sensitivity for model $\mathcal{B}$ and a common squark mass
(black dashed line). The green band shows the uncertainty implied by
varying the scale in
the range $m/2 \le \mu\le 2m$, where $\mu=\mu_{\text{F}}=\mu_{\text{R}}$ 
is the factorization and renormalization scale and $m$ is the averaged mass 
of the produced sparticles.
The error from scale dependence translates into
uncertainties of roughly $\pm20\GEV$ to $\pm40\GEV$
in the squark and gluino masses.
The lower right panel of figure \ref{fig:scale} shows the relative
error $|\sigma_{m/2}-\sigma_{2m}|/2\sigma_{m}$ of each production
channel as a function of the ratio $\mgo/\msq$, which
is varied along the sensitivity limit shown in the left panel.
The upper right panel of figure \ref{fig:scale} shows the ratio of the respective
cross section to the
total cross section, again
along the sensitivity curve. The average relative error is around
0.2 and is lowest in the region of a dominant $\sq\sq$ channel.
The $\go\go$ production involves very large uncertainties for large $\mgo/\msq$.
However, in this region the cross section for $\go\go$
production is very small.

\section{Stopped staus}\label{sec:stopped}

So far we considered the signal of staus as heavy stable charged 
particles passing the detectors. As we already mentioned in section 
\ref{sec:lowbeta} very slow staus can lose their kinetic energy 
completely and get trapped inside the detector. These staus will 
then decay into the LSP inside the detector.
This causes a signal 
that can be recorded with dedicated trigger algorithms
\cite{Asai:2009ka,Pinfold:2010aq,Graham:2011ah}.
The decay of the stau reveals very interesting information about the
LSP and possibly even about the origin of SUSY breaking. 
Once the mass of the 
stau is known\footnote{%
The mass is determined already at the stage of discovery by
measuring the momentum and the velocity
\cite{Aad:2011hz,CMS-PAS-EXO-11-022}.
}
one can determine the LSP mass from reconstructed 2-body decays. 
Furthermore, for a gravitino LSP
an unequivocal prediction of supergravity can be tested, 
namely the proportionality of the stau lifetime
\beq \label{eq:taustau}
\tau_\stau\simeq\frac{48\pi m_{3/2}^2 M_{\text{P}}^2}{\mstau^5}\left(1-\frac{m_{3/2}^2}{\mstau^2}\right)^{-4}
\eeq
to the Planck mass squared \cite{Buchmuller:2004rq,Feng:2004gn}. 
In scenarios with an axino LSP, stau decays may provide insights into the
Peccei-Quinn sector~\cite{Brandenburg:2005he}. 

In this section we will estimate the prospects for the
LHC to observe decays of stopped staus by
computing the number of staus that are expected to be stopped inside the
detectors in the framework of our simplified models.%
\footnote{Since a detailed 
study of the LHC's potential for measuring the stau 
lifetime and LSP mass requires a precise detector simulation 
taking into account details of the LHC operating schedule,
we leave this to the 
experimental collaborations.
Combining our results with those of \cite{Pinfold:2010aq}
it may be possible to estimate
the prospects for measuring the lifetime in the simplified models.}
The mean range $R$ of a charged particle 
(defined as the average thickness 
of absorber material the particle is 
capable of traversing) grows linearly with the particle's mass.
Its dependence on $\gamma\beta$
is nearly linear in the double logarithmic 
plot \cite{0954-3899-37-7A-075021} 
in the region of interest $\gamma\beta<1$. In fact,
\beq
\label{eq:range}
\log_{10}\left(\frac{R/\text{g}\,\text{cm}^{-2}}{m/\!\GEV}\right) = c_1 + c_2\log_{10}\left(\gamma\beta\right)
\eeq
approximates $R(\gamma\beta)$ to a precision better than 1\%. For iron
$c_1\simeq2.16$ and $c_2\simeq3.32$.
From (\ref{eq:range}) we determine the maximal velocity of a stau with mass
$\mstau$ to be expected to stop inside the detector, 
$\beta_{\max}^{R}(\mstau)$.
Conservatively, we assume a maximal range of 
$R_{\max}=2400\,\text{g}\,\text{cm}^{-2}$.
This is a conservative estimation of the thickness
of a CMS-like detector in the central region.\footnote{%
We took into account 20 layers of silicon (2.33$\,$g$/$cm$^3$), each $0.5\,$mm thick;
1 layer of ECAL crystal (8.28$\,$g$/$cm$^3$), $23\,$cm thick;
16 layers of HCAL brass (8.53$\,$g$/$cm$^3$), each around $6\,$cm thick;
1 magnet of NbTi (5.6$\,$g$/$cm$^3$), $31.2\,$cm thick;
3 layers of iron yokes (7.87$\,$g$/$cm$^3$), 30, 63 and $63\,$cm thick 
\cite{Chatrchyan:2008aa}.
}
\begin{figure}
\centering
\setlength{\unitlength}{1\textwidth}
\begin{picture}(0.85,0.4)
 \put(0.0,0.0){
  \put(0.04,0.035){\includegraphics[scale=1.1]{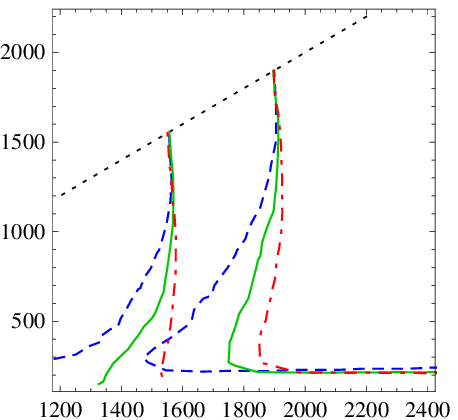}}   
  \put(0.15,0.0){\footnotesize $\mgo=\msq\; [ \GEV \, ] $}
  \put(0.0,0.14){\rotatebox{90}{\footnotesize $\mstau\;[\GEV\,]$}}
  \put(0.09,0.316){\footnotesize $\,pp\,(8\TEV)$}
  \put(0.09,0.286){\footnotesize $\int\!\mathcal{L}=16\ifb$}
  \put(0.23,0.26){\scriptsize \color{gray} 1}
  \put(0.14,0.21){\scriptsize \color{gray} 10}
  }
 \put(0.43,0){
  \put(0.04,0.035){\includegraphics[scale=1.1]{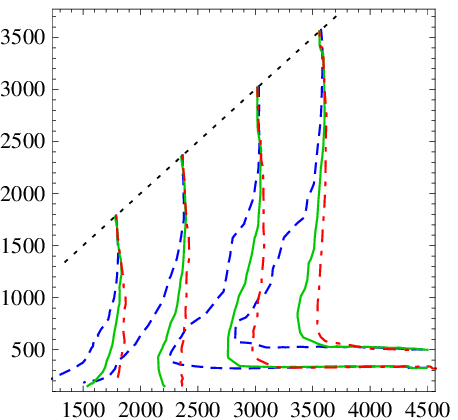}}   
  \put(0.15,0.0){\footnotesize $\mgo=\msq\; [ \GEV \, ] $}
  \put(0.0,0.14){\rotatebox{90}{\footnotesize $\mstau\;[\GEV\,]$}}
  \put(0.09,0.316){\footnotesize $\,pp\,(14\TEV)$}
  \put(0.09,0.286){\footnotesize $\int\!\mathcal{L}=300\ifb$}
  \put(0.27,0.283){\scriptsize\color{gray} 1}
  \put(0.21,0.24){\scriptsize \color{gray} 10}
  \put(0.144,0.186){\scriptsize \color{gray} 100}
  \put(0.084,0.137){\scriptsize \color{gray} 1000}
  }
\end{picture}
\caption{Number of events that contain at least one stau which is stopped inside
the detector for the simplified models $\mathcal{A}$ (blue dashed), 
$\mathcal{B}$ (green solid) and $\mathcal{C}$ (red dot-dashed). 
The iso-event-number curves are displayed in the $\mgo$-$\mstau$ plane
where we have chosen the common squark mass scenario and $\msq=\mgo$.
}
\label{fig:stopped}
\end{figure}

Figure \ref{fig:stopped} shows the expected number of events that provide
at least one stau which is expected to stop inside the detector, i.e.,
with $\beta<\beta_{\max}^{R}(\mstau)$.
According to the discussion in section \ref{sec:decay}, for large mass gaps
$\mlcp-\mstau$ staus are expected to be
considerably slower in model $\mathcal{C}$ than in model $\mathcal{A}$. 
As a consequence, the numbers of stopped staus typically differ by an order of magnitude.

At the $8\TEV$ LHC only a few stopping events are expected in the
mass region of interest.
Thus, most likely it will not be possible to study the properties of stau decays 
in detail. The high-luminosity $14\TEV$ run, however,  will 
provide reasonable numbers of stopped staus in scenarios that
lie within the discovery reach of the full $8\TEV$ dataset or the early 
$14\TEV$ dataset. Consequently, such scenarios are expected to be accessible 
in sufficient detail to determine
the stau lifetime.  Determining the LSP mass seems feasible as well,
unless $m_\text{LSP} \lesssim 0.1\,\mstau$, in which case the
mass determination becomes extremely difficult because it requires
a very precise measurement of the tau recoil energy. 

A more detailed study of stau decays including the measurement of the 
spin of the LSP via reconstructed 3-body decays
\cite{Buchmuller:2004rq,Brandenburg:2005he}
may only be possible at an $e^+e^-$ collider
with a dedicated detector environment, see e.g.\
\cite{Hamaguchi:2004df,Feng:2004yi,Martyn:2006as,Cakir:2007xa}.

\section{Conclusions}

We have studied the potential of the LHC to discover or exclude scenarios with a 
long-lived stau (or another charged slepton) in a simplified-model approach.
The production of colored sparticles is likely to be the dominant production mode 
at the LHC\@.
SUSY events are characterized by muon-like particles that leave the detector and 
can be slow. In addition, hard jets can be expected in some cases.
We have defined $2\times3$ discrete simplified models covering the limiting cases
concerning the production and the decay, respectively. Each model contains only 
three free parameters, $\mstau$, $\mgo$ and either the common squark mass 
$\msq$ or $\mstop$. We have also included the Drell-Yan production of stau pairs.

Due to the very prominent signature of long-lived staus, exclusion and discovery 
take place on the basis of a very few events. Consequently, a discovery could be 
established in a rather short period of time. Due to the direct DY contribution, 
regions where both staus are typically too fast to be identified (where SM particle 
radiation would provide a more significant signal) are not present.
For the same reason, no very specific cuts are required, enabling a 
model-independent analysis.  In other words, it is possible to cover the whole 
parameter space with a small number of selection criteria that yield both a high 
signal efficiency and a very good background rejection.  In almost the whole
parameter space, we have found a signal efficiency around $50\%$.  Even
in the most challenging region, the efficiency does not drop below roughly
$20\%$.  This shows that in the long-lived stau scenario no regions exist where 
the theory effectively hides from observation.  This is a striking difference to the 
neutralino LSP scenario, where both compressed spectra and very stretched 
ones are very hard to observe. 

If all squarks have a common mass the most conservative projected limits for 
$\int\!\mathcal{L}=16\ifb$ at the $8\TEV$ LHC ($\int\!\mathcal{L}=300\ifb$ at $14\TEV$)
are $\msq\gtrsim1.6\TEV$ ($\msq\gtrsim3.3\TEV$). If a stop is significantly lighter 
than the other squarks the corresponding limits become $\mstop\gtrsim950\GEV$ 
($\mstop\gtrsim2\TEV$).
In both cases the gluino is expected to be either excluded or discovered
up to a mass of $\mgo\simeq1.4\TEV$ ($\mgo\simeq 2.6\TEV$).
Intermediate cases may be estimated by interpolating the results according to the
discussion in section \ref{sec:prod}. Already the data collected so far should allow 
one to place lower limits above a TeV on both $\msq$ and $\mgo$ even in the 
most challenging scenario. The actual experimental searches might achieve even 
better sensitivities, since we chose quite conservative assumptions concerning 
background rejection and detector efficiencies.

Staus stopped in the detectors could provide intriguing possibilities to test 
supergravity or to gain insights into the SUSY breaking mechanism.
We have computed the number of stopped stau events in the framework
of the simplified models. 
Especially for very compressed spectra, the LHC provides a very good 
environment to measure the stau lifetime.

\subsection*{Acknowledgements}
We would like to thank Sergei Bobrovskyi, Jim Brooke, Silja Brensing, 
Giacomo Bruno, Jie Chen, James Hirschauer, Kolja Kaschube, Tomas Kasemets, 
Boris Panes, Loic Quer\-ten\-mont, Peter Schleper and Daniel Wiesler for very 
helpful discussions.
Very special thanks go to Dao Thi Nhung and Sabine Kraml for generously
providing their \textsc{SDECAY} extension for three-body lepton decays.
This work was supported by the German Science Foundation (DFG) via the
Junior Research Group `SUSY Phenomenology' within the Collaborative
Research Center 676 `Particles, Strings and the Early Universe'.

\phantomsection 
\bibliographystyle{../utphys}
\bibliography{../staus}

\providecommand{\href}[2]{#2}\begingroup\raggedright\begin{thebibliography}{10}

\bibitem{Djouadi:1998di}
MSSM Working Group, A.~Djouadi {\em et al.}, ``{The Minimal Supersymmetric
  Standard Model: Group Summary Report}'',
  \href{http://arxiv.org/abs/hep-ph/9901246}{{\tt arXiv:hep-ph/9901246}}.

\bibitem{Berger:2008cq}
C.~F. Berger, J.~S. Gainer, J.~L. Hewett, and T.~G. Rizzo, ``{Supersymmetry
  Without Prejudice}'',
  \href{http://dx.doi.org/10.1088/1126-6708/2009/02/023}{{\em JHEP} {\bf 0902}
  (2009)  023},
\href{http://arxiv.org/abs/0812.0980}{{\tt arXiv:0812.0980 [hep-ph]}}.

\bibitem{Alwall:2008ag}
J.~Alwall, P.~Schuster, and N.~Toro, ``{Simplified Models for a First
  Characterization of New Physics at the LHC}'',
  \href{http://dx.doi.org/10.1103/PhysRevD.79.075020}{{\em Phys. Rev.} {\bf
  D79} (2009)  075020},
\href{http://arxiv.org/abs/0810.3921}{{\tt arXiv:0810.3921 [hep-ph]}}.

\bibitem{Alves:2011wf}
LHC New Physics Working Group, D.~Alves {\em et al.}, ``{Simplified Models for
  LHC New Physics Searches}'',
\href{http://arxiv.org/abs/1105.2838}{{\tt arXiv:1105.2838 [hep-ph]}}.

\bibitem{1997NuPhB.492...51B}
W.~Beenakker, R.~H{\"o}pker, M.~Spira, and P.~Zerwas, ``{Squark and gluino
  production at hadron colliders}'',
  \href{http://dx.doi.org/10.1016/S0550-3213(97)80027-2}{{\em Nucl. Phys.} {\bf
  B492} (1997)  51--103}, \href{http://arxiv.org/abs/hep-ph/9610490}{{\tt
  arXiv:hep-ph/9610490}}.
  \url{http://www.thphys.uni-heidelberg.de/~plehn/index.php?show=prospino}.

\bibitem{Konar:2010bi}
P.~Konar, K.~T. Matchev, M.~Park, and G.~K. Sarangi, ``{How to look for
  supersymmetry under the lamppost at the LHC}'',
  \href{http://dx.doi.org/10.1103/PhysRevLett.105.221801}{{\em Phys. Rev.
  Lett.} {\bf 105} (2010)  221801},
\href{http://arxiv.org/abs/1008.2483}{{\tt arXiv:1008.2483 [hep-ph]}}.

\bibitem{Horn:2009zx}
C.~Horn, ``{A Bottom-Up Approach to SUSY Analyses}'',
  \href{http://dx.doi.org/10.1088/0954-3899/36/10/105005}{{\em J. Phys.} {\bf
  G36} (2009)  105005},
\href{http://arxiv.org/abs/0905.4497}{{\tt arXiv:0905.4497 [hep-ex]}}.

\bibitem{Gavin:2010az}
R.~Gavin, Y.~Li, F.~Petriello, and S.~Quackenbush, ``{FEWZ 2.0: A code for
  hadronic $Z$ production at next-to-next-to-leading order}'',
  \href{http://dx.doi.org/10.1016/j.cpc.2011.06.008}{{\em Comput. Phys.
  Commun.} {\bf 182} (2011)  2388--2403},
  \href{http://arxiv.org/abs/1011.3540}{{\tt arXiv:1011.3540 [hep-ph]}}.
\url{http://gate.hep.anl.gov/fpetriello/FEWZ.html}.

\bibitem{Campbell:2011bn}
J.~M. Campbell, R.~Ellis, and C.~Williams, ``{Vector boson pair production at
  the LHC}'', \href{http://dx.doi.org/10.1007/JHEP07(2011)018}{{\em JHEP} {\bf
  1107} (2011)  018},
\href{http://arxiv.org/abs/1105.0020}{{\tt arXiv:1105.0020 [hep-ph]}}.

\bibitem{Campbell:2004ch}
J.~M. Campbell, R.~K. Ellis, and F.~Tramontano, ``{Single top production and
  decay at next-to-leading order}'',
  \href{http://dx.doi.org/10.1103/PhysRevD.70.094012}{{\em Phys. Rev.} {\bf
  D70} (2004)  094012},
\href{http://arxiv.org/abs/hep-ph/0408158}{{\tt arXiv:hep-ph/0408158}}.

\bibitem{Campbell:2005bb}
J.~M. Campbell and F.~Tramontano, ``{Next-to-leading order corrections to W t
  production and decay}'',
  \href{http://dx.doi.org/10.1016/j.nuclphysb.2005.08.015}{{\em Nucl. Phys.}
  {\bf B726} (2005)  109--130},
\href{http://arxiv.org/abs/hep-ph/0506289}{{\tt arXiv:hep-ph/0506289}}.

\bibitem{Campbell:2008hh}
J.~M. Campbell, R.~K. Ellis, F.~Febres~Cordero, F.~Maltoni, L.~Reina,
  D.~Wackeroth, and S.~Willenbrock, ``{Associated production of a $W$ boson and
  one $b$ jet}'', \href{http://dx.doi.org/10.1103/PhysRevD.79.034023}{{\em
  Phys. Rev.} {\bf D79} (2009)  034023},
\href{http://arxiv.org/abs/0809.3003}{{\tt arXiv:0809.3003 [hep-ph]}}.

\bibitem{Caola:2011pz}
J.~Campbell, F.~Caola, F.~Febres~Cordero, L.~Reina, and D.~Wackeroth, ``{NLO
  QCD predictions for $W+1$ jet and $W+2$ jet production with at least one $b$
  jet at the 7 TeV LHC}'',
  \href{http://dx.doi.org/10.1103/PhysRevD.86.034021}{{\em Phys.Rev.} {\bf D86}
  (2012)  034021},
\href{http://arxiv.org/abs/1107.3714}{{\tt arXiv:1107.3714 [hep-ph]}}.

\bibitem{MCFMweb}
J.~M. Campbell and R.~Ellis, {\em MCFM -- Monte Carlo for FeMtobarn processes}.
\newblock \url{http://mcfm.fnal.gov/}.

\bibitem{Alwall:2011uj}
J.~Alwall, M.~Herquet, F.~Maltoni, O.~Mattelaer, and T.~Stelzer, ``{MadGraph 5:
  Going Beyond}'', \href{http://dx.doi.org/10.1007/JHEP06(2011)128}{{\em JHEP}
  {\bf 1106} (2011)  128}, \href{http://arxiv.org/abs/1106.0522}{{\tt
  arXiv:1106.0522 [hep-ph]}}.

\bibitem{sjostrand-2006-0605}
T.~Sj{\"o}strand, S.~Mrenna, and P.~Skands, ``{PYTHIA 6.4 physics and
  manual}'', \href{http://dx.doi.org/10.1088/1126-6708/2006/05/026}{{\em JHEP}
  {\bf 0605} (2006)  026}, \href{http://arxiv.org/abs/hep-ph/0603175}{{\tt
  arXiv:hep-ph/0603175}}.

\bibitem{alwall-2008-53}
J.~Alwall, S.~H{\"o}che, F.~Krauss, N.~Lavesson, L.~L{\"o}nnblad, F.~Maltoni,
  M.~L. Mangano, M.~Moretti, C.~G. Papadopoulos, F.~Piccinini, S.~Schumann,
  M.~Treccani, J.~Winter, and M.~Worek, ``Comparative study of various
  algorithms for the merging of parton showers and matrix elements in hadronic
  collisions'', \href{http://dx.doi.org/10.1140/epjc/s10052-007-0490-5}{{\em
  Eur. Phys. J.} {\bf C53} (2008)  473--500},
  \href{http://arxiv.org/abs/0706.2569}{{\tt arXiv:0706.2569 [hep-ph]}}.

\bibitem{Pumplin:2002vw}
J.~Pumplin, D.~R. Stump, J.~Huston, H.~L. Lai, P.~Nadolsky, and W.~K. Tung,
  ``{New generation of parton distributions with uncertainties from global QCD
  analysis}'', \href{http://dx.doi.org/10.1088/1126-6708/2002/07/012}{{\em
  JHEP} {\bf 0207} (2002)  012},
  \href{http://arxiv.org/abs/hep-ph/0201195}{{\tt arXiv:hep-ph/0201195}}.

\bibitem{Ovyn:2009tx}
S.~Ovyn, X.~Rouby, and V.~Lemaitre, ``{DELPHES, a framework for fast simulation
  of a generic collider experiment}'',
  \href{http://arxiv.org/abs/0903.2225}{{\tt arXiv:0903.2225 [hep-ph]}}.
  \url{http://www.fynu.ucl.ac.be/users/s.ovyn/Delphes/download.html}.

\bibitem{Cacciari:2008gp}
M.~Cacciari, G.~P. Salam, and G.~Soyez, ``{The anti-$k_t$ jet clustering
  algorithm}'', \href{http://dx.doi.org/10.1088/1126-6708/2008/04/063}{{\em
  JHEP} {\bf 0804} (2008)  063},
\href{http://arxiv.org/abs/0802.1189}{{\tt arXiv:0802.1189 [hep-ph]}}.

\bibitem{Aad:2011hz}
ATLAS Collaboration, G.~Aad {\em et al.}, ``{Search for heavy long-lived
  charged particles with the ATLAS detector in $pp$ collisions at $\sqrt{s} =
  7$ TeV}'', \href{http://dx.doi.org/10.1016/j.physletb.2011.08.042}{{\em Phys.
  Lett.} {\bf B703} (2011)  428--446},
\href{http://arxiv.org/abs/1106.4495}{{\tt arXiv:1106.4495 [hep-ex]}}.

\bibitem{CMS-PAS-EXO-11-022}
{{CMS} Collaboration}, ``{Search for Heavy Stable Charged Particles}'',
  CMS-PAS-EXO-11-022, CERN, Geneva, Jul, 2011.
\newblock \url{http://cds.cern.ch/record/1370057}.

\bibitem{CMS-PAS-EXO-10-004}
{{CMS} Collaboration}, ``{Search for Heavy Stable Charged Particles in $pp$
  collisions at $\sqrt{s}=7$ TeV}'',  CMS-PAS-EXO-10-004, CERN, Geneva, Jul,
  2010.
\newblock \url{http://cds.cern.ch/record/1280690}.

\bibitem{CMS-PAS-EXO-08-003}
{{CMS} Collaboration}, ``Search for heavy stable charged particles with 100
  pb$^{-1}$ and 1 fb$^{-1}$ in the {CMS} experiment'',  CMS-PAS-EXO-08-003,
  CERN, Geneva, Feb, 2009.
\newblock \url{http://cdsweb.cern.ch/record/1152570}.

\bibitem{0954-3899-37-7A-075021}
Particle Data Group, K.~Nakamura {\em et al.}, ``{Review of particle
  physics}'',
\href{http://dx.doi.org/10.1088/0954-3899/37/7A/075021}{{\em J. Phys.} {\bf
  G37} (2010)  075021}.

\bibitem{Muhlleitner:2003vg}
M.~M{\"u}hlleitner, A.~Djouadi, and Y.~Mambrini, ``{SDECAY: A Fortran code for
  the decays of the supersymmetric particles in the MSSM}'',
  \href{http://dx.doi.org/10.1016/j.cpc.2005.01.012}{{\em Comput. Phys.
  Commun.} {\bf 168} (2005)  46--70},
\href{http://arxiv.org/abs/hep-ph/0311167}{{\tt arXiv:hep-ph/0311167}}.

\bibitem{Kilian:2007gr}
W.~Kilian, T.~Ohl, and J.~Reuter, ``{WHIZARD: Simulating Multi-Particle
  Processes at LHC and ILC}'',
  \href{http://dx.doi.org/10.1140/epjc/s10052-011-1742-y}{{\em Eur. Phys. J.}
  {\bf C71} (2011)  1742}, \href{http://arxiv.org/abs/0708.4233}{{\tt
  arXiv:0708.4233 [hep-ph]}}.
\url{http://projects.hepforge.org/whizard/}.

\bibitem{KoljasThesis}
K.~Kaschube, {\em {Search for Stable Stau Production at the LHC}}.
\newblock PhD thesis, University of Hamburg, 2011.
\newblock
\url{http://inspirehep.net/record/940823}.
\newblock

\bibitem{Heisig:2011dr}
J.~Heisig and J.~Kersten, ``{Production of long-lived staus in the Drell-Yan
  process}'', \href{http://dx.doi.org/10.1103/PhysRevD.84.115009}{{\em Phys.
  Rev.} {\bf D84} (2011)  115009},
\href{http://arxiv.org/abs/1106.0764}{{\tt arXiv:1106.0764 [hep-ph]}}.

\bibitem{Lindert:2011td}
J.~M. Lindert, F.~D. Steffen, and M.~K. Trenkel, ``{Direct stau production at
  hadron colliders in cosmologically motivated scenarios}'',
  \href{http://dx.doi.org/10.1007/JHEP08(2011)151}{{\em JHEP} {\bf 1108} (2011)
   151},
\href{http://arxiv.org/abs/1106.4005}{{\tt arXiv:1106.4005 [hep-ph]}}.

\bibitem{Kraml:2007sx}
S.~Kraml and D.~Nhung, ``{Three-body decays of sleptons in models with
  non-universal Higgs masses}'',
  \href{http://dx.doi.org/10.1088/1126-6708/2008/02/061}{{\em JHEP} {\bf 0802}
  (2008)  061},
\href{http://arxiv.org/abs/0712.1986}{{\tt arXiv:0712.1986 [hep-ph]}}.

\bibitem{Beenakker:2011fu}
W.~Beenakker, S.~Brensing, M.~Kr{\"a}mer, A.~Kulesza, E.~Laenen, {\em et al.},
  ``{Squark and gluino hadroproduction}'',
  \href{http://dx.doi.org/10.1142/S0217751X11053560}{{\em Int. J. Mod. Phys.}
  {\bf A26} (2011)  2637--2664},
\href{http://arxiv.org/abs/1105.1110}{{\tt arXiv:1105.1110 [hep-ph]}}.

\bibitem{Asai:2009ka}
S.~Asai, K.~Hamaguchi, and S.~Shirai, ``{Measuring lifetimes of long-lived
  charged massive particles stopped in LHC detectors}'',
  \href{http://dx.doi.org/10.1103/PhysRevLett.103.141803}{{\em Phys. Rev.
  Lett.} {\bf 103} (2009)  141803},
\href{http://arxiv.org/abs/0902.3754}{{\tt arXiv:0902.3754 [hep-ph]}}.

\bibitem{Pinfold:2010aq}
J.~Pinfold and L.~Sibley, ``{Measuring the Lifetime of Trapped Sleptons Using
  the General Purpose LHC Detectors}'',
  \href{http://dx.doi.org/10.1103/PhysRevD.83.035021}{{\em Phys. Rev.} {\bf
  D83} (2011)  035021},
\href{http://arxiv.org/abs/1006.3293}{{\tt arXiv:1006.3293 [hep-ph]}}.

\bibitem{Graham:2011ah}
P.~W. Graham, K.~Howe, S.~Rajendran, and D.~Stolarski, ``{New Measurements with
  Stopped Particles at the LHC}'',
  \href{http://dx.doi.org/10.1103/PhysRevD.86.034020}{{\em Phys. Rev.} {\bf
  D86} (2012)  034020},
\href{http://arxiv.org/abs/1111.4176}{{\tt arXiv:1111.4176 [hep-ph]}}.

\bibitem{Buchmuller:2004rq}
W.~Buchm{\"u}ller, K.~Hamaguchi, M.~Ratz, and T.~Yanagida, ``Supergravity at
  colliders'', \href{http://dx.doi.org/10.1016/j.physletb.2004.03.016}{{\em
  Phys. Lett.} {\bf B588} (2004)  90--98},
\href{http://arxiv.org/abs/hep-ph/0402179}{{\tt arXiv:hep-ph/0402179}}.

\bibitem{Feng:2004gn}
J.~L. Feng, A.~Rajaraman, and F.~Takayama, ``{Probing gravitational
  interactions of elementary particles}'',
  \href{http://dx.doi.org/10.1142/S0218271804006474}{{\em Int. J. Mod. Phys.}
  {\bf D13} (2004)  2355--2359},
\href{http://arxiv.org/abs/hep-th/0405248}{{\tt arXiv:hep-th/0405248}}.

\bibitem{Brandenburg:2005he}
A.~Brandenburg, L.~Covi, K.~Hamaguchi, L.~Roszkowski, and F.~D. Steffen,
  ``Signatures of axinos and gravitinos at colliders'',
  \href{http://dx.doi.org/10.1016/j.physletb.2005.04.072}{{\em Phys. Lett.}
  {\bf B617} (2005)  99--111},
\href{http://arxiv.org/abs/hep-ph/0501287}{{\tt arXiv:hep-ph/0501287}}.

\bibitem{Chatrchyan:2008aa}
CMS Collaboration, S.~Chatrchyan {\em et al.}, ``{The CMS experiment at the
  CERN LHC}'',
\href{http://dx.doi.org/10.1088/1748-0221/3/08/S08004}{{\em JINST} {\bf 3}
  (2008)  S08004}.

\bibitem{Hamaguchi:2004df}
K.~Hamaguchi, Y.~Kuno, T.~Nakaya, and M.~M. Nojiri, ``{A Study of late decaying
  charged particles at future colliders}'',
  \href{http://dx.doi.org/10.1103/PhysRevD.70.115007}{{\em Phys. Rev.} {\bf
  D70} (2004)  115007},
\href{http://arxiv.org/abs/hep-ph/0409248}{{\tt arXiv:hep-ph/0409248}}.

\bibitem{Feng:2004yi}
J.~L. Feng and B.~T. Smith, ``{Slepton trapping at the large hadron and
  international linear colliders}'',
  \href{http://dx.doi.org/10.1103/PhysRevD.71.015004,
  10.1103/PhysRevD.71.019904}{{\em Phys. Rev.} {\bf D71} (2005)  015004},
\href{http://arxiv.org/abs/hep-ph/0409278}{{\tt arXiv:hep-ph/0409278}}.

\bibitem{Martyn:2006as}
H.-U. Martyn, ``{Detecting metastable staus and gravitinos at the ILC}'',
  \href{http://dx.doi.org/10.1140/epjc/s2006-02630-7}{{\em Eur. Phys. J.} {\bf
  C48} (2006)  15--24},
\href{http://arxiv.org/abs/hep-ph/0605257}{{\tt arXiv:hep-ph/0605257}}.

\bibitem{Cakir:2007xa}
O.~\c{C}ak{\i}r, {\.I}.~T. \c{C}ak{\i}r, J.~R. Ellis, and Z.~K{\i}rca, ``{Study
  of Metastable Staus at Linear Colliders}'',
\href{http://arxiv.org/abs/hep-ph/0703121}{{\tt arXiv:hep-ph/0703121}}.

\end{thebibliography}\endgroup

\end{document}